\documentclass[%
 aip,
 apl,
 amsmath,amssymb,
 reprint,%
 floatfix,
]{revtex4-2}

\usepackage{graphicx}
\usepackage{dcolumn}
\usepackage{bm}
\usepackage{silence}
\usepackage{siunitx}
\usepackage[utf8]{inputenc}
\usepackage[T1]{fontenc}
\usepackage{mathptmx}
\usepackage{etoolbox}
\usepackage{xcolor} 
\usepackage[colorlinks=true, allcolors=blue]{hyperref}
\usepackage[nameinlink]{cleveref}

\usepackage{float}
\usepackage{booktabs}
\DeclareSIUnit\mbar{mbar}

\makeatletter
\def\@email#1#2{%
 \endgroup
 \patchcmd{\titleblock@produce}
  {\frontmatter@RRAPformat}
  {\frontmatter@RRAPformat{\produce@RRAP{*#1\href{mailto:#2}{#2}}}\frontmatter@RRAPformat}
  {}{}
}%
\makeatother

\begin{document}

\preprint{AIP/123-QED}


\title[APL Photonics]{Soft X-Ray Detection Down to the Nitrogen K-Edge Using a Pixelated iLGAD Sensor}


\author{Shuqi Li}
\email{shuqi.li@psi.ch}
\affiliation{Center for Photon Science, Paul Scherrer Institut, Villigen PSI, Switzerland}


\author{Anna Bergamaschi}
\affiliation{Center for Photon Science, Paul Scherrer Institut, Villigen PSI, Switzerland}

\author{Ashish Bisht}
\affiliation{Fondazione Bruno Kessler, Trento, Italy}

\author{Maurizio Boscardin}
\affiliation{Fondazione Bruno Kessler, Trento, Italy}

\author{Martin Brückner}
\affiliation{Center for Photon Science, Paul Scherrer Institut, Villigen PSI, Switzerland}

\author{Maria Carulla}
\affiliation{Center for Photon Science, Paul Scherrer Institut, Villigen PSI, Switzerland}

\author{Tomáš Celko}
\affiliation{Center for Photon Science, Paul Scherrer Institut, Villigen PSI, Switzerland}

\author{Roberto Dinapoli}
\affiliation{Center for Photon Science, Paul Scherrer Institut, Villigen PSI, Switzerland}

\author{Simon Ebner}
\affiliation{Center for Photon Science, Paul Scherrer Institut, Villigen PSI, Switzerland}

\author{Simone Emiliani}
\affiliation{Center for Photon Science, Paul Scherrer Institut, Villigen PSI, Switzerland}

\author{Khalil Ferjaoui}
\affiliation{Center for Photon Science, Paul Scherrer Institut, Villigen PSI, Switzerland}

\author{Erik Fröjdh}
\affiliation{Center for Photon Science, Paul Scherrer Institut, Villigen PSI, Switzerland}

\author{Viveka Gautam}
\affiliation{Center for Photon Science, Paul Scherrer Institut, Villigen PSI, Switzerland}

\author{Dominic Greiffenberg}
\affiliation{Center for Photon Science, Paul Scherrer Institut, Villigen PSI, Switzerland}

\author{Shqipe Hasanaj}
\affiliation{Center for Photon Science, Paul Scherrer Institut, Villigen PSI, Switzerland}

\author{Viktoria Hinger}
\affiliation{Center for Photon Science, Paul Scherrer Institut, Villigen PSI, Switzerland}

\author{Marius Hürst}
\affiliation{Center for Photon Science, Paul Scherrer Institut, Villigen PSI, Switzerland}

\author{Vadym Kedych}
\affiliation{Center for Photon Science, Paul Scherrer Institut, Villigen PSI, Switzerland}

\author{Thomas King}
\affiliation{Center for Photon Science, Paul Scherrer Institut, Villigen PSI, Switzerland}

\author{Antonio Liguori}
\affiliation{Dipartimento Interateneo di Fisica, University of Bari "Aldo Moro", Bari, Italy}

\author{Carlos Lopez-Cuenca}
\affiliation{Center for Photon Science, Paul Scherrer Institut, Villigen PSI, Switzerland}

\author{Leonid Lunin}
\affiliation{Center for Photon Science, Paul Scherrer Institut, Villigen PSI, Switzerland}

\author{Alice Mazzoleni}
\affiliation{Center for Photon Science, Paul Scherrer Institut, Villigen PSI, Switzerland}

\author{Davide Mezza}
\affiliation{Center for Photon Science, Paul Scherrer Institut, Villigen PSI, Switzerland}

\author{Scharon Moses}
\affiliation{Center for Photon Science, Paul Scherrer Institut, Villigen PSI, Switzerland}

\author{Konstantinos Moustakas}
\affiliation{Center for Photon Science, Paul Scherrer Institut, Villigen PSI, Switzerland}

\author{Aldo Mozzanica}
\affiliation{Center for Photon Science, Paul Scherrer Institut, Villigen PSI, Switzerland}

\author{Jonathan Mulvey}
\affiliation{Center for Photon Science, Paul Scherrer Institut, Villigen PSI, Switzerland}

\author{Martin Müller}
\affiliation{Center for Photon Science, Paul Scherrer Institut, Villigen PSI, Switzerland}

\author{Giovanni Paternoster}
\affiliation{Fondazione Bruno Kessler, Trento, Italy}

\author{Christian Ruder}
\affiliation{Center for Photon Science, Paul Scherrer Institut, Villigen PSI, Switzerland}

\author{Bernd Schmitt}
\affiliation{Center for Photon Science, Paul Scherrer Institut, Villigen PSI, Switzerland}

\author{Saverio Silletta}
\affiliation{Center for Photon Science, Paul Scherrer Institut, Villigen PSI, Switzerland}

\author{Dhanya Thattil}
\affiliation{Center for Photon Science, Paul Scherrer Institut, Villigen PSI, Switzerland}


\author{Matteo Centis Vignali}
\affiliation{Fondazione Bruno Kessler, Trento, Italy}


\author{Xiangyu Xie}
\affiliation{Center for Photon Science, Paul Scherrer Institut, Villigen PSI, Switzerland}

\author{Jiaguo Zhang}
\affiliation{Center for Photon Science, Paul Scherrer Institut, Villigen PSI, Switzerland}

\author{Sinuo Zhang}
\affiliation{Center for Photon Science, Paul Scherrer Institut, Villigen PSI, Switzerland}

\date{\today}


\begin{abstract}

Hybrid pixel detectors represent the state-of-the-art technology for hard X-ray experiments at synchrotron light sources and free-electron lasers, owing to their high frame rates, extended dynamic range, and scalable detection areas. However, their application in the soft X-ray regime remains bottlenecked by the challenge of developing a sensor that can achieve high quantum efficiency and an sufficient signal-to-noise ratio at low photon energies. Here, we present inverse low-gain avalanche diode (iLGAD) sensors integrated with a thin entrance window, demonstrating single-photon resolution down to 390 eV (below the Nitrogen K-edge) with a signal-to-noise ratio of 5. Two gain layer designs, standard gain layer and shallow gain layer, were systematically investigated to evaluate their spectral response as a function of photon energy and operating temperature. A double-peak feature is observed in the energy spectrum, arising from the dependence of charge multiplication on photon absorption depth. Furthermore, a phenomenological spectral fitting model was established, enabling the quantitative extraction of the effective $\mathrm{n}^+$ layer thickness and gain layer depth. Beyond providing a fundamental understanding of iLGAD charge multiplication mechanisms and characterization results in the soft X-ray regime, these findings lay the foundation for developing large-area, high-frame-rate hybrid pixel detectors for soft X-ray. This advancement directly benefits soft X-ray photon science applications, including high-resolution ptychography, spectroscopy, and resonant scattering at synchrotron and FEL beamlines.

\end{abstract}





\maketitle


\section{Introduction}\label{sec1}
Soft X-ray radiation (250~eV--2~keV) provides a unique probe for accessing the K-edges of low-$Z$ elements for structural and chemical analysis; in particular, the ``water window'' (284--543~eV) between the carbon and oxygen K-edges offers high intrinsic contrast for biological imaging. Furthermore, this energy regime covers the L-edges of $3d$ transition metals for probing magnetic and electronic states, as well as the M-edges of rare-earth elements. To achieve these scientific goals, a wide range of techniques operate in this energy regime, ranging from spectroscopic and scattering probes, including resonant X-ray diffraction \cite{Scagnoli2011} and resonant inelastic X-ray scattering \cite{Monney2022}, to high-resolution lensless imaging, most notably ptychography \cite{Hitchcock2015}. The successful implementation of these techniques relies heavily on 2D pixelated detectors. Driven by recent advances in accelerator-based light sources, particularly free-electron lasers like SwissFEL and EuXFEL alongside diffraction-limited storage rings like MAX IV and SLS 2.0, light sources are reaching unprecedented levels of brilliance and spatial coherence. Concurrently, the emergence of high-harmonic generation (HHG) technology has made tabletop soft X-ray sources viable. Together, these developments across large-scale facilities and laboratory platforms offer more opportunities across science and industry, while posing significant challenges to current detector technologies in the soft X-ray regimes.

Hybrid pixel detectors (HPDs), consisting of a sensor and a readout chip interconnected via bump-bonding, are widely used in the tender and hard X-ray regimes (> 2 keV). At the Paul Scherrer Institut (PSI), we have developed high-performance HPDs—such as JUNGFRAU~\cite{Mozzanica2014, JungmannSmith2016, Leonarski2020}, MÖNCH~\cite{Dinapoli2014, Cartier2016, Ramilli2017}, and EIGER~\cite{Tinti2015, Tinti2018, Butcher2023}—which are deployed at facilities worldwide. Their widespread adoption originates from their high dynamic range (up to $10^4$~photons/pixel at 12~keV), fast frame rates ($>2$~kHz), and modular scalability for large-area coverage. However, their application below \SI{2}{keV} is fundamentally challenged by the low quantum efficiency (QE) and poor signal-to-noise ratio (SNR) of the silicon sensors in HPDs. In contrast, conventional detectors commonly used in the soft X-ray regime, such as charge-coupled devices (CCDs) and complementary metal-oxide-semiconductor (CMOS) imagers, often suffer from restricted active areas, low frame rates, and limited dynamic range. As a result, they increasingly struggle to meet the high readout speed and wide dynamic range demanded by next-generation soft X-ray experiments. Therefore, extending the application of HPDs to the soft X-ray regime offers a promising route to overcome these limitations.

In the soft X-ray regime, shallow photon absorption near the sensor surface reduces the QE, while the small charge generated by soft X-rays is comparable to electronic noise, leading to a poor SNR. Crucial to mitigating these limitations in HPDs below \SI{2}{keV} is the optimization of the sensor, which absorbs incoming X-ray photons and transfers the generated signal to the readout chip. Therefore, to enable HPDs to operate in the soft X-ray regime, PSI has collaborated with Fondazione Bruno Kessler (FBK) to develop an optimized sensor: inverse low-gain avalanche diode (iLGAD) sensors incorporating a thin entrance window (TEW)~\cite{Zhang2022}. LGAD technology introduces a doping-engineered gain layer that amplifies the photo-generated signal via impact ionization, with an internal gain typically ranging from 5 to 20. This signal amplification effectively raises the signal well above the noise floor and improves the overall SNR, thereby enabling single-photon discrimination with Poisson-limited statistics, provided that the detector system is not dominated by shot noise in the sensor. Crucially, this internal gain does not compromise the advantages of HPDs~\cite{Zhang2022, Hinger2024}. Simultaneously, the integration of TEW technology minimizes shallow surface absorption, allowing low-energy photons down to the EUV regime to penetrate into the active silicon volume.


Previously, we have reported the QE of the developed TEW technology elsewhere~\cite{Carulla2023, Carulla2024}.
In this work, we optimize the pixelated iLGAD sensors with two different gain layer designs and characterize the spectral response using the charge-integrating MÖNCH readout chip under synchrotron radiation down to the nitrogen K-edge (\qty{390}{eV}), achieving a SNR of 5 at \qty{400}{eV}. The measurements revealed a double-peak structure in the spectral response, which is attributed to the distinct charge-multiplication processes initiated by electrons and holes, resulting from photon absorption at varying depths within the device. This additional signal, previously not observed in the photon-counting detector~\cite{Baruffaldi2025}, contributes to an enhancement of the overall detection efficiency, a direct outcome of the implemented TEW technology. The temperature effects on the multiplication factor in the iLGAD are also presented. These promising results establish a clear pathway to extend the performance of hybrid detectors into the soft X-ray regime, thereby overcoming the current detector bottleneck in soft X-ray science.

\section{Methods}\label{sec2}

\subsection{iLGAD sensor optimization and design}

Standard LGAD sensors incorporate a dedicated gain layer beneath each pixel's electrode on the front side; however, this architecture leads to a reduced fill factor and a loss of signal multiplication in the inter-pixel regions, especially for fine-pitch pixel devices. In contrast, iLGAD sensors feature a single, continuous gain layer on the backside that spans the entire pixel array. This design achieves a uniform spatial response with a 100\% fill factor, enabling high-precision spatial interpolation. Furthermore, integrating the TEW with a $n^{+}$-doping layer onto the backside is compatible with the iLGAD structure.

The electric field profile within the sensor is designed to produce controllable impact ionization in a linear gain regime while avoiding avalanche breakdown. This electric field distribution is primarily achieved by tuning doping profiles of the $n^{+}$ and $p^{+}$ layers, as well as the applied bias voltage. In thick substrates, such as the \qty{275}{\um} silicon used in this work, the influence of external bias on the localized gain region is limited, rendering the implanted $n^{+}$ and $p^{+}$ doping profiles the primary determinants of the gain mechanism.

\begin{figure*}[tb] 
  \centering
  \begin{minipage}[t]{0.45\textwidth}
    \raggedright \textbf{(a)} \vspace{-2pt}\\
    \centering
    \includegraphics[width=\linewidth, height=0.32\textheight, keepaspectratio]{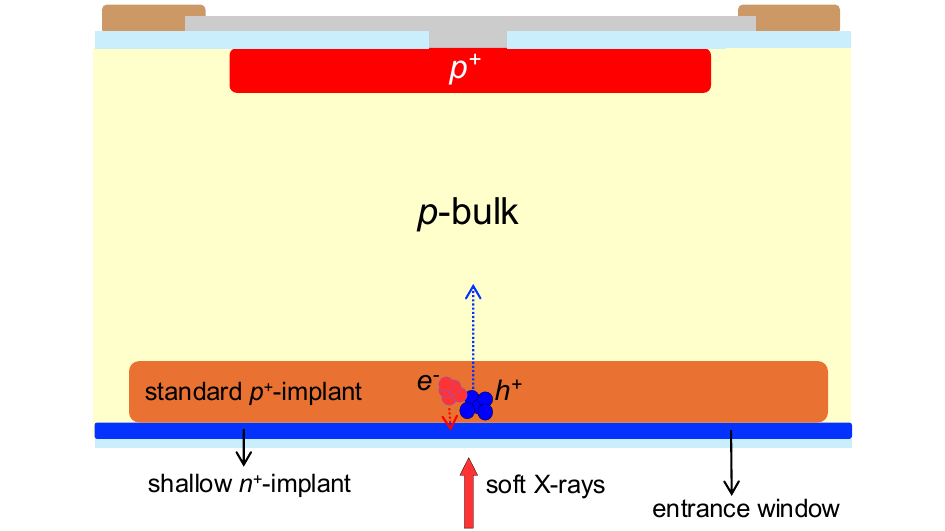}%
    \label{fig:sumup_a} 
  \end{minipage}%
  \hfill
  \begin{minipage}[t]{0.45\textwidth}
    \raggedright \textbf{(b)} \vspace{-2pt}\\
    \centering
    \includegraphics[width=\linewidth, height=0.32\textheight, keepaspectratio]{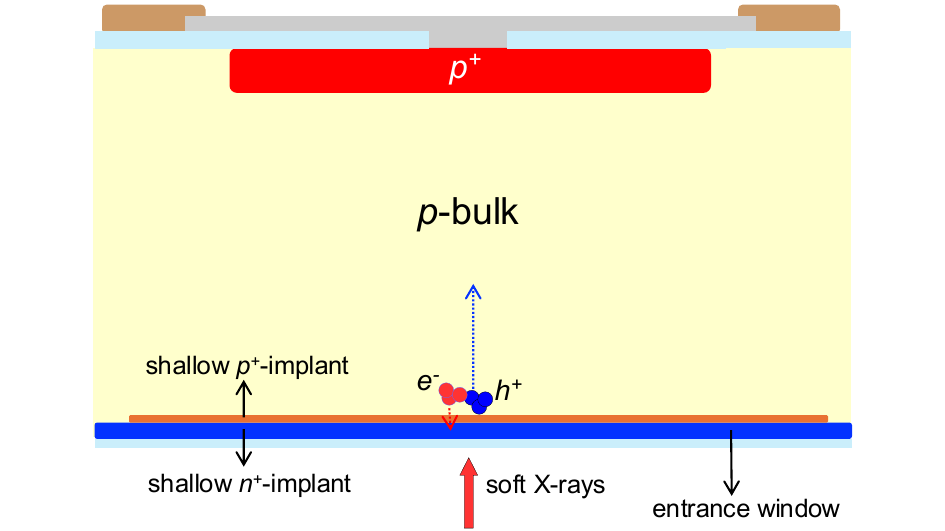}%
    \label{fig:sumup_b}
  \end{minipage}

  \vspace{0.15cm} 

  \begin{minipage}[t]{0.42\textwidth}
    \raggedright \textbf{(c)} \vspace{-2pt}\\
    \centering
    \includegraphics[width=\linewidth, height=0.32\textheight, keepaspectratio]{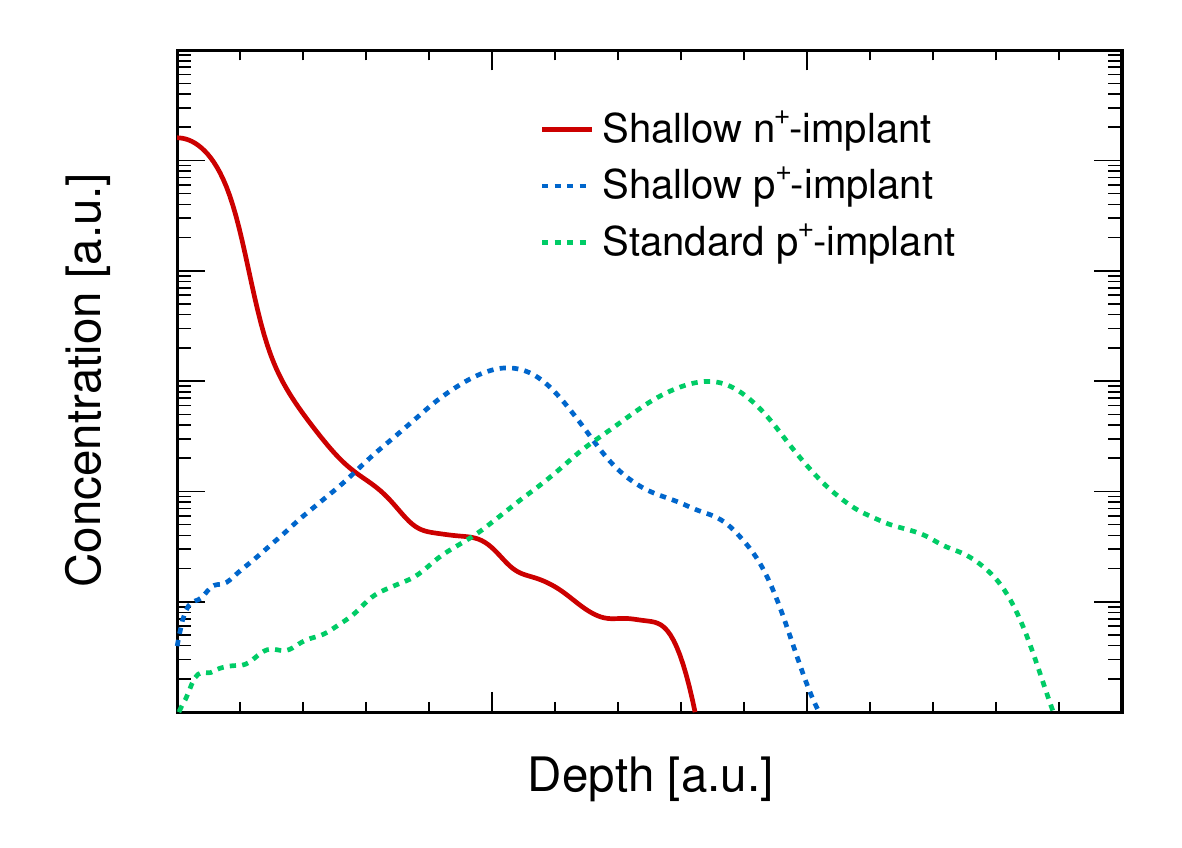}%
    \label{fig:sumup_c}
  \end{minipage}%
  \hfill
  \begin{minipage}[t]{0.42\textwidth}
    \raggedright \textbf{(d)} \vspace{-2pt}\\
    \centering
    \includegraphics[width=\linewidth, height=0.32\textheight, keepaspectratio]{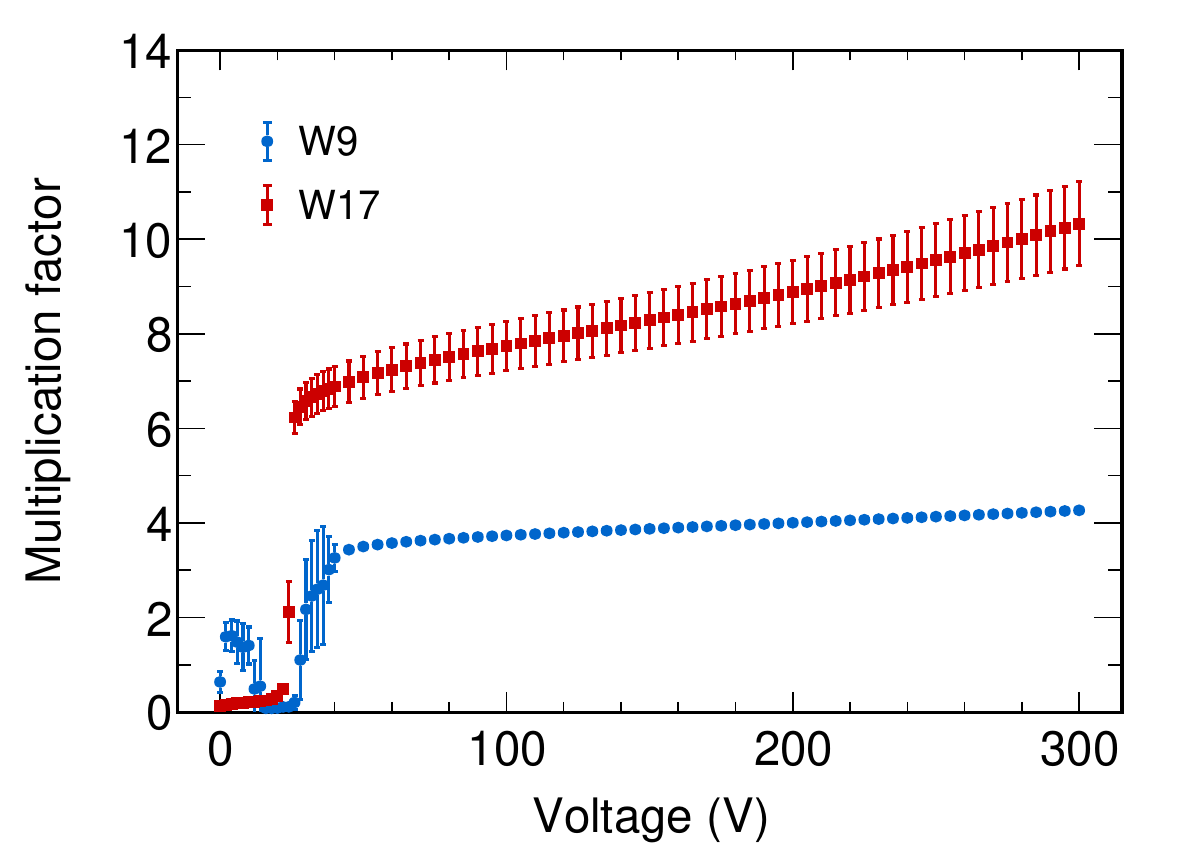}%
    \label{fig:sumup_d}
  \end{minipage}

  \vspace{-2pt} 
  \caption{Optimization strategies for iLGADs: (a)~Cross-section of the standard gain-layer design and (b)~cross-section of the shallow gain-layer design. (c)~Comparison of simulated $n^+$ and $p$-gain doping profiles for both designs. (d)~Measured multiplication factors for the standard and shallow designs under \SI{800}{\nano\meter} LED illumination at room temperature.}
  \label{fig:fig_iLGAD_opt}
\end{figure*}

Incident backside-illuminated X-rays are absorbed at varying depths in silicon following the Beer-Lambert law, where the absorption probability decays exponentially with depth, governed by the photon-energy-dependent attenuation length. When photons are absorbed shallowly within the $n^+$ layer (i.e., before reaching the gain layer), the photo-generated holes drift toward the gain layer under the internal electric field to initiate multiplication. Conversely, for absorption occurring beyond the gain layer, electrons initiate the multiplication process. Because the impact ionization coefficient of electrons in silicon is a few times larger than that of holes, electron-initiated multiplication produces more secondary charge carriers and thus larger signal amplitudes. Consequently, maximizing the fraction of electron-initiated events is desirable, which can be achieved by minimizing the depths of the $n^+$ layer and $p$-gain layer. Accordingly, the first iLGAD batch developed in this work was fabricated with both standard and shallow gain layers for comparison as shown in Fig.~\ref{fig:fig_iLGAD_opt}(a) and (b).


The corresponding doping profiles are shown in Fig.~\ref{fig:fig_iLGAD_opt}(c), which were simulated through the process simulator of Synopsys TCAD alongside the electric field profiles. Both designs utilize an identical $n^{+}$ profile optimized for the TEW process; however, their $p$-gain profiles differ significantly. In the shallow design, the peak of the $p$-gain layer is located at approximately half the depth of the standard design. To compensate for this narrower high-field region, a higher peak $p$-doping concentration was implemented. 

Experimentally, the multiplication factors were determined by comparing the photocurrents of diodes with and without the gain layer under \SI{800}{\nano\meter} LED illumination. The multiplication factor is defined as the ratio of the photocurrent in the gain-layer device to that in the control (no-gain) device. Because \SI{800}{\nano\meter} light has an absorption depth of $\sim$\SI{11.8}{\micro\meter} in silicon---well beyond the gain layer---the measured amplification shown in Fig.~\ref{fig:fig_iLGAD_opt}(d) is predominantly governed by electron-initiated multiplication. The multiplication factor of the shallow gain layer is approximately half that of the standard design. These LED-derived results align well with soft X-ray measurements performed at a synchrotron source \cite{Liguori2023}. The nominal gains at 300 V are $\sim$10.2 for the standard and $\sim$4.2 for the shallow gain layer designs, respectively. This different is not intrinsically due to the different profile depth, but also to a mismatch in the doping concentration used resulting in a gain lower then predicted by the simulation for the shallow gain layer.

\begin{figure}[ht]
\includegraphics[width=0.4\textwidth]{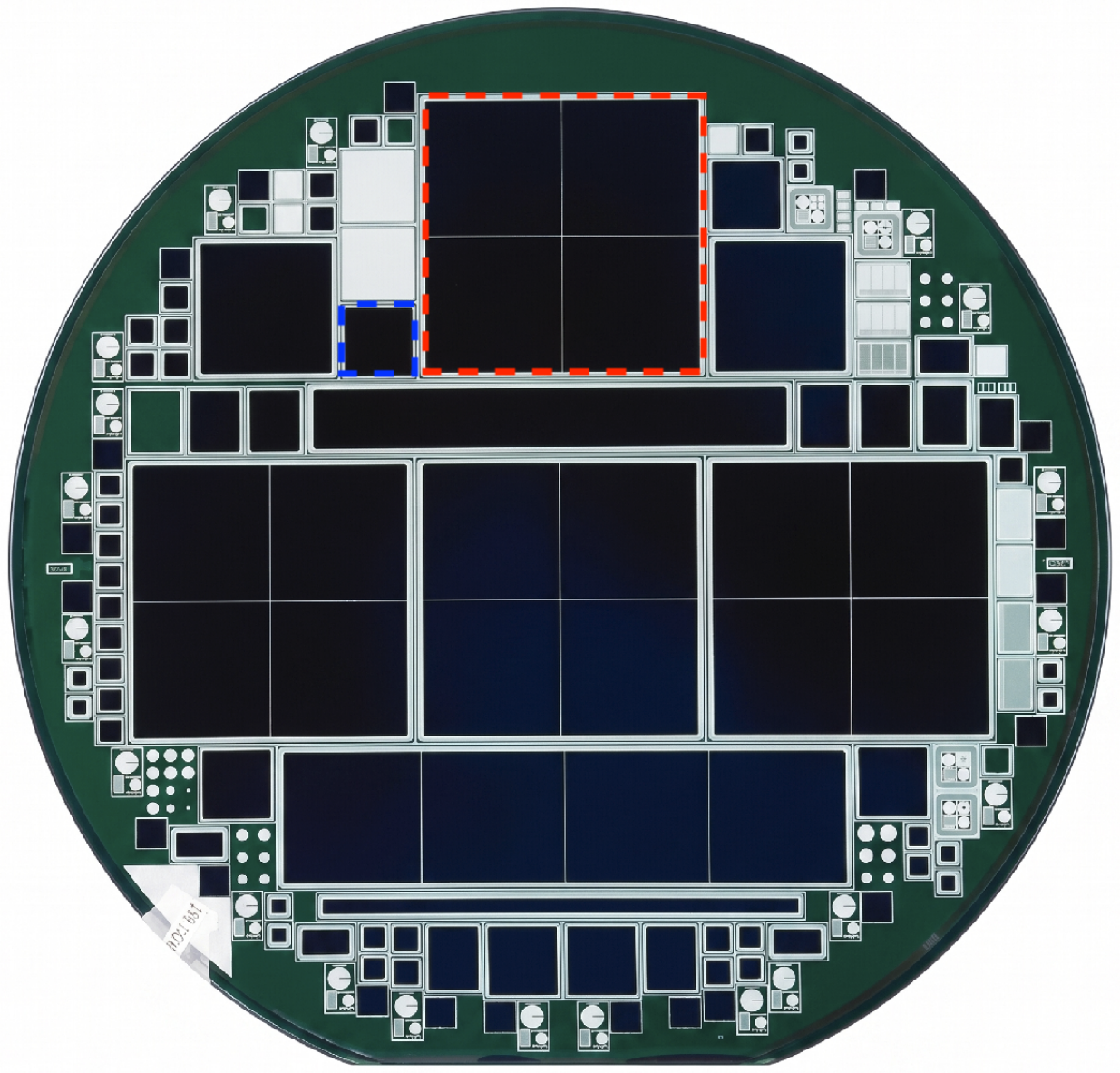}
\caption{\label{fig_wafer_TEW_iLGAD} Fabricated iLGAD wafer with a thin entrance window: The sensor with 4 cm $\times$ 4 cm and 75 $\mu$m pixel pitch marked in the dashed red line is the largest iLGAD sensor produced so far to the best knowledge of the authors. The 1 cm $\times$ 1 cm sensors with 25 $\mu$m pixel pitch, which are used in this study, are marked in the dashed blue rectangle.}
\end{figure}

Standard and shallow iLGAD sensors incorporating the optimized TEW process were fabricated on 6-inch wafers at FBK (Fig.~\ref{fig_wafer_TEW_iLGAD}). The wafer features several sensor designs, including a large-area \qtyproduct{4x4}{\cm} device with a \SI{75}{\micro\meter} pixel pitch, which represents the largest iLGAD sensor fabricated to date. It also includes \qtyproduct{1x1}{\cm} sensors with a \SI{25}{\micro\meter} pixel pitch, which is the focus of this study, alongside process control monitor (PCM) diodes for wafer-level qualification.

\subsection{iLGAD sensor characterization}\label{subsec32}

\begin{figure}[ht]
\centering
\includegraphics[width=0.42\textwidth]{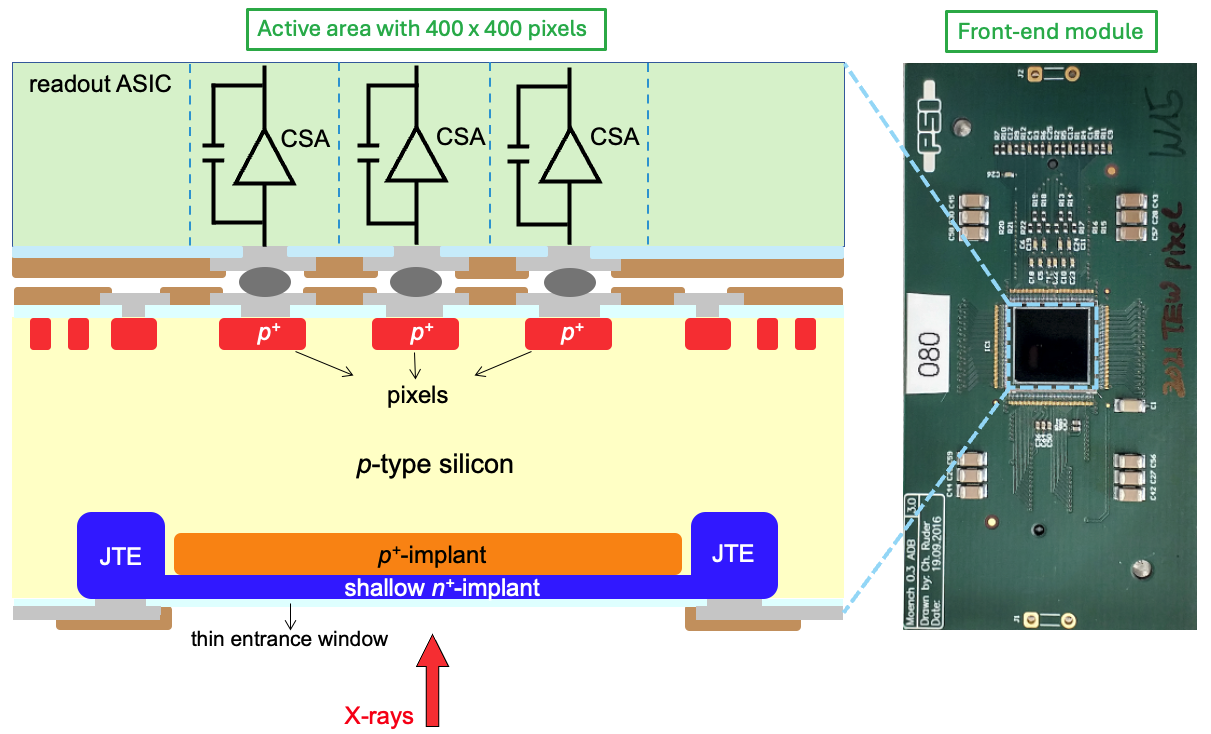}
\caption{Cross-sectional view and photograph of the front-end iLGAD-MÖNCH module.}
\label{fig:fig_FE_module_iLGAD}
\end{figure}

The characterized iLGAD sensors feature a \qty{25}{\um} pixel pitch and is bump-bonded to a MÖNCH readout ASIC \cite{Dinapoli2014, Ramilli2017}, comprising an array of \numproduct{400 x 400} pixels that covers an active area of \qty{1}{\cm\squared}. As shown in Fig.~\ref{fig:fig_FE_module_iLGAD}, the pixel matrix is surrounded by a CCR and a multi-guard ring structure to prevent an early breakdown. MÖNCH is a charge-integrating readout chip featuring a charge-sensitive amplifier (CSA) and a correlated double sampling (CDS) circuit for low-frequency noise suppression, with its differential analog outputs digitized by off-chip analog-to-digital converters (ADCs) on the readout board. Following pedestal subtraction from the dark frames, the detector output, expressed in ADC units (ADCu), scales linearly with the charge deposited by incident X-ray photons.

\subsubsection{Measurement set-up}

\begin{figure}[ht]
\centering
\includegraphics[width=\columnwidth]{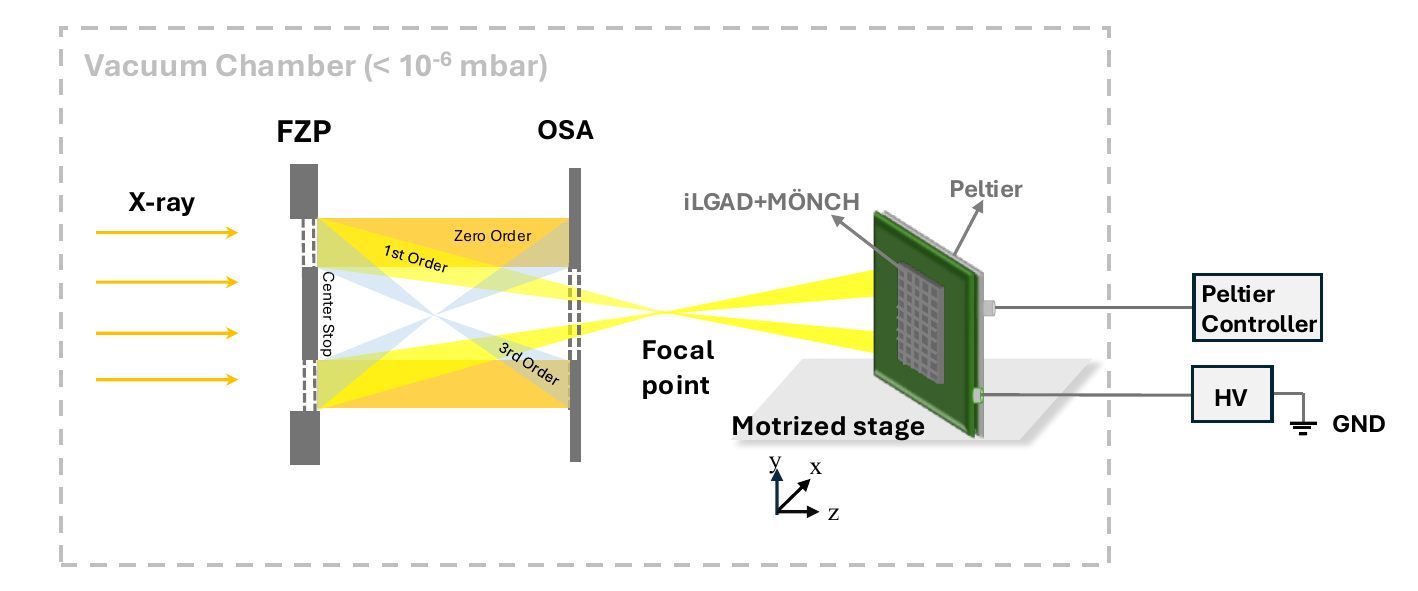}
\caption{Experimental setup at the SoftiMAX beamline of MAX IV. After exiting the slits, the X-ray beam was focused and spatial-filtered by a Fresnel zone plate (FZP) and an order sorting aperture (OSA) before impinging on the iLGAD sensor. All optical components were mounted on a rail inside the vacuum chamber.}
\label{fig:fig_setup_spectra}
\end{figure}


The beam test setup, shown in Fig.~\ref{fig:fig_setup_spectra}, was conducted at the SoftiMAX beamline of MAX IV, which provides tunable soft X-rays ranging from \qtyrange{280}{2500}{\eV}. The iLGAD-MÖNCH module was mounted inside a vacuum chamber ($< 10^{-6}$~mbar), facing the incident soft X-ray beam. The beam was focused using a FZP and passed through an OSA to reject higher-order harmonics. The detector was placed $\sim$\,\qty{20}{\cm} downstream from the focal plane, illuminating an area larger than the beam spot size by diffusing the beam. At each photon energy, the exit slit dimensions were adjusted to maintain an optimal occupancy (<1\%), ensuring single-photon detection within the \qty{10}{\us} exposure window. Prior to each data acquisition run, a total of 1000 pedestal frames were collected without X-ray illumination to correct the energy spectra and evaluate the noise. To evaluate the temperature dependence of the gain and noise, the module was cooled using a Peltier element, which allowed for adjusting and monitoring the temperature at the module backplane. During the beam test, the temperature was varied from \qtyrange{-50}{20}{\celsius}, with the sensor biased at \qty{300}{\V}, well above its depletion voltage of $\sim$\,\qty{160}{\V}.


\subsubsection{Spectral characterization and fitting}
\label{sec:fitting}
Due to the shallow absorption depth of soft X-rays, most photons are absorbed near the backside of the sensor. The generated charge clouds drift through the \qty{275}{\um}-thick silicon substrate with a drift time of $\sim$\qty{7}{\ns}, diffusing laterally to a spatial width of $\sim$\qty{11}{\um} (Full Width of Half Maximum). This lateral diffusion leads to charge sharing among adjacent \qty{25}{\micro\meter}-pitch pixels. To accurately reconstruct the photon energy, a cluster-finding algorithm~\cite{Cartier2014}, implemented in the AARE data analysis framework~\cite{Frojdh2025}, is applied to each pedestal-subtracted frame. Pixels are selected if their signals exceed $3\sigma_{\text{noise}}$, and those with the highest charge among their neighbors are identified as cluster centers. The noise root-mean-square (r.m.s.), denoted as $\sigma_{\text{noise}}$, represents the noise level in ADC units (ADCu). This value was derived from the dark current measurements for each individual pixel. In this work, $2 \times 2$ pixel clusters are used to reconstruct the total deposited charge. A cluster originating from a true single photon is selected only if its total charge exceeds $2 \times 5\sigma_{\text{noise}}$, where the factor of 2 accounts for the cluster size. Finally, the charges of the valid clusters are filled into a histogram to form the energy spectrum.

\begin{figure*}[ht]
\centering
(a)\includegraphics[width=0.42\textwidth]{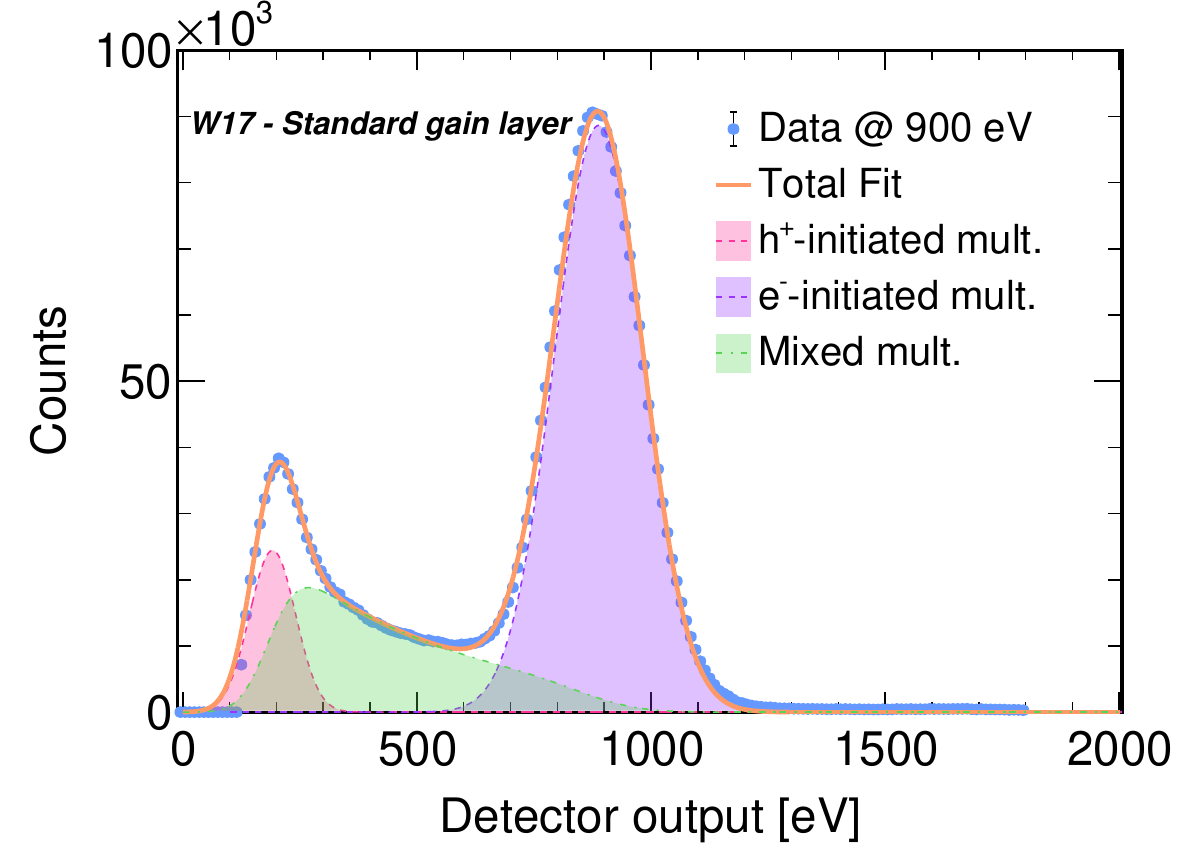}
(b)\includegraphics[width=0.42\textwidth]{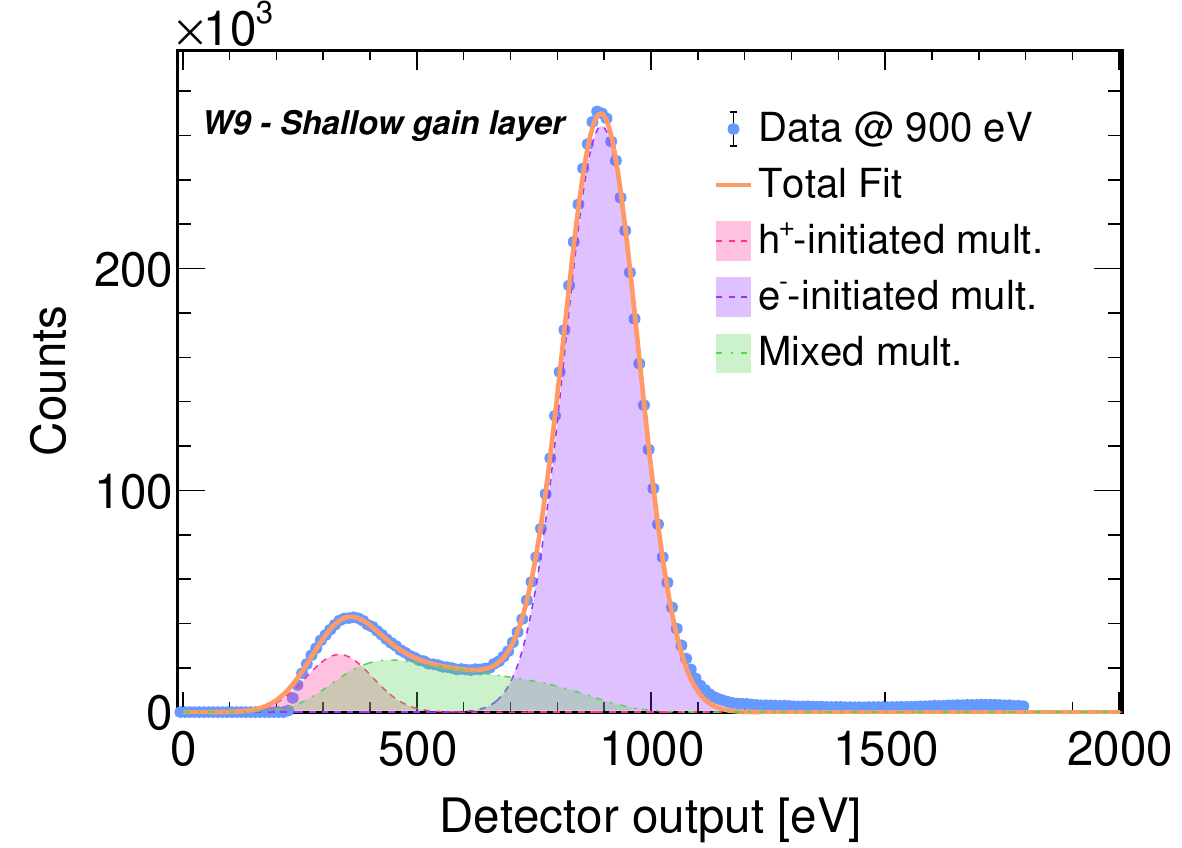}
\caption{Measured spectral response at \qty{900}{\eV} for (a) the standard gain design and (b) the shallow gain design, together with spectral fittings decomposed into electron-initiated, hole-initiated, and co-initiated (electron-hole) multiplication components.
}
\label{fig:fig_fitting_procedure}
\end{figure*}

As shown in Fig.~\ref{fig:fig_fitting_procedure}, the iLGAD spectral response measured with 900 eV soft X-ray photons exhibits a characteristic double-peak structure. The peak position at the high detector output has been calibrated to the photon energy. This distinctive feature arises from the sensor's internal structure and the variation in photon absorption depth governed by the Beer–Lambert law. When photons are absorbed within the $n^+$ layer (the blue region in Fig.~\ref{fig:fig_FE_module_iLGAD}), the generated holes drift toward the gain layer and trigger impact ionization, giving rise to the hole-initiated multiplication peak. Conversely, when photons are absorbed deeper in the $p$-bulk region beyond the gain layer, the primary electrons drift into the gain layer, resulting in electron-initiated multiplication. Because the impact ionization coefficient of electrons is higher than that of holes \cite{VANOVERSTRAETEN1970583, 10114953}, the signal generated by electron-initiated multiplication is larger, thus creating a second peak at a higher output position in the spectrum. Finally, for photons absorbed within the gain layer, both electrons and holes contribute to the multiplication process, forming the continuous region between the two primary peaks. Because the shallow gain design features a thinner gain layer than the standard design, a higher probability of photon absorption beyond the gain layer is expected at the same photon energy. Consequently, the shallow gain design benefits from a higher proportion of electron-initiated multiplication, as compared in Fig.~\ref{fig:fig_fitting_procedure}.

According to the Beer-Lambert law, the probability of photon absorption at different depths depends on the attenuation length of photons in silicon at the corresponding energy. Lower photon energies lead to a higher proportion of hole-initiated multiplication. To quantitatively analyze the variation in the proportions of these three spectral components, a fitting model was developed which is formulated as the superposition of two discrete Gaussian peaks corresponding to pure hole- and electron-initiated multiplication, together with a smoothed, exponentially decaying continuum representing mixed multiplication within the gain layer. Therefore, the total spectrum $f_{\text{total}}(x)$ is described as:

\begin{equation}
f_{\text{total}}(x) = f_{\text{h}}(x) + f_{\text{e}}(x) + f_{\text{mixed}}(x)
\end{equation}

\noindent where the pure hole- and electron-initiated multiplication peaks $f_{\text{h}}(x)$ and $f_{\text{e}}(x)$ are parameterized as:


\begin{align}
f_{\text{h}}(x) &= A_{\text{h}} \exp\left( -\frac{(x - \mu_{\text{h}})^2}{2\sigma_{\text{h}}^2} \right) \\
f_{\text{e}}(x) &= A_{\text{e}} \exp\left( -\frac{(x - \mu_{\text{e}})^2}{2\sigma_{\text{e}}^2} \right)
\end{align}
where $\mu_{\text{h}}$ ($\mu_{\text{e}}$) and $\sigma_{\text{h}}$ ($\sigma_{\text{e}}$) denote the mean positions and Gaussian noise widths of the hole- (electron-) initiated peaks, and $A_{\text{h}}$ ($A_{\text{e}}$) represents their relative amplitudes.

The continuum component $f_{\text{mixed}}(x)$ accounts for a linear gain variation between the two regions and arises from continuous photon absorption within the gain layer according to the Beer--Lambert law. Specifically, each absorption depth yields a signal centered at a position corresponding to the local multiplication gain, convolved with Gaussian electronic noise ($\sigma_{\text{h}}$ and $\sigma_{\text{e}}$ at the boundaries of the hole- and electron-initiated peaks, respectively). Integrating these noise-broadened responses across the gain region yields a smooth box profile bounded between $\mu_{\text{h}}$ and $\mu_{\text{e}}$ via error functions. Furthermore, as the photon flux attenuates exponentially with depth following the Beer--Lambert law, the continuum height decreases exponentially from $\mu_{\text{h}}$ toward $\mu_{\text{e}}$ with a decay constant $\lambda$. Therefore, $f_{\text{mixed}}(x)$ is described by:

\begin{equation}
\frac{A_{\text{box}}}{2} \left[ \text{erf}\left( \frac{x - \mu_{\text{h}}}{\sqrt{2}\sigma_{\text{h}}} \right) - \text{erf}\left( \frac{x - \mu_{\text{e}}}{\sqrt{2}\sigma_{\text{e}}} \right) \right] \exp\left( -\frac{x - \mu_{\text{h}}}{\lambda} \right)
\end{equation}

\noindent Here,  $A_{\text{box}}$ is the amplitude of the mixed continuum at the lower boundary, and $\lambda$ represents the effective decay constant corresponding to the photon absorption depth profile. 

Overall, the total fitting function involves eight free parameters. The fitting was performed using a binned weighted least-squares ($\chi^2$) minimization implemented via the Minuit2 minimizer in ROOT. Point-by-point statistical errors were modeled by assuming Poisson statistics ($\sigma_i = \sqrt{y_i}$). To ensure robust convergence, a two-stage fitting strategy was adopted: preliminary segmented fits were conducted on localized sub-regions (Gaussian functions for the primary peaks and an exponential function for the intermediate continuum) to extract reliable initial parameter estimates, which were then passed to the global full-spectrum optimization. Fig.~\ref{fig:fig_fitting_procedure} illustrates the three fitted components for both standard and shallow gain LGAD designs at \qty{900}{\eV}, showing good agreement with the measured data.


\section{Results}\label{sec3}

The iLGAD sensors with two different gain layers were characterized in terms of spectral response, temperature dependence, and SNR. In addition, the depth of the n$^{+}$-implant below the entrance window and the depth of the gain layer have been determined from the measurement. This section presents the experimental results for these sensors using soft X-rays from synchrotron radiation sources and the determined depths for the implanted region close to the surface.

\subsubsection{Energy Calibration}
\label{sec:energy_calibration}

\begin{figure}[ht]
    \centering
    \includegraphics[width=0.42\textwidth]{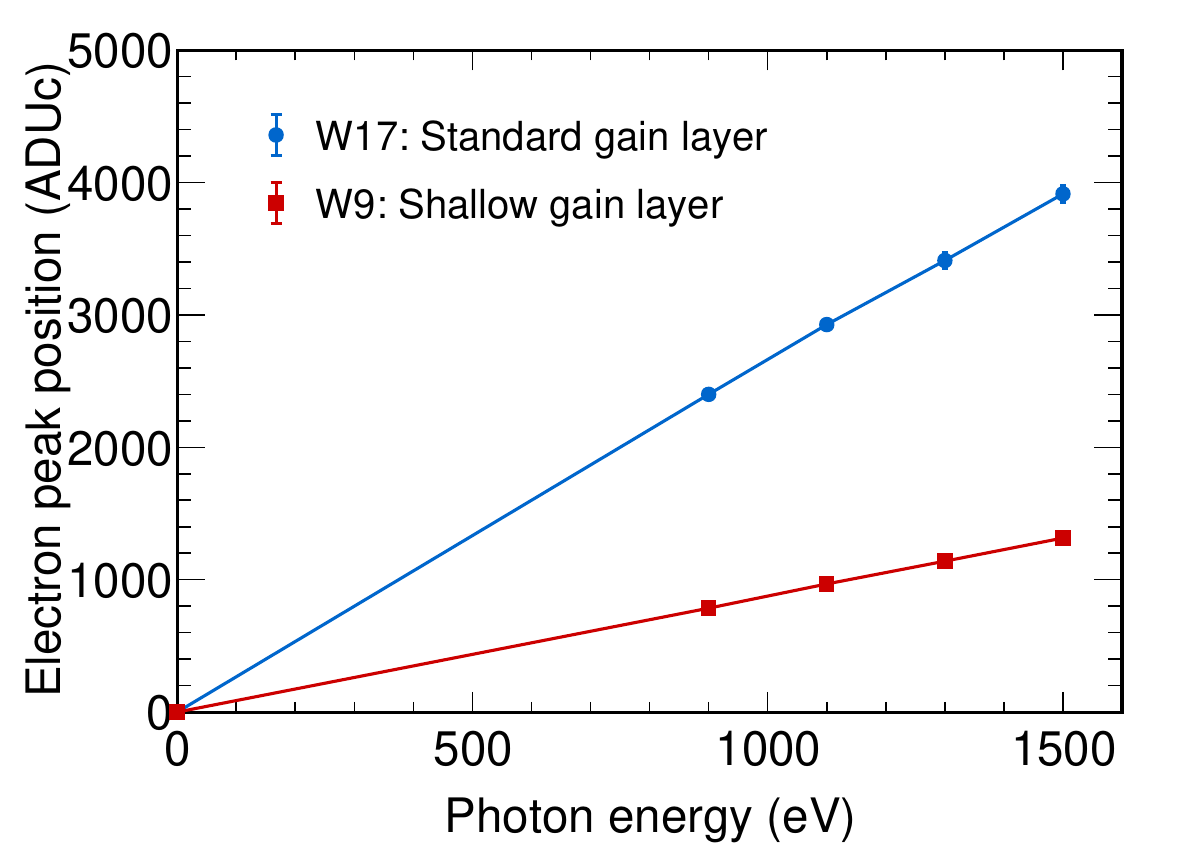}
    \caption{\label{fig:energy_calibration} Measured electron peak position in ADUc as a function of photon energy for the W17 (standard gain layer) and W9 (shallow gain layer) iLGAD designs.}
\end{figure}

To calibrate the detector response to soft X-ray photon energies, the electron-initiated multiplication peak positions (in $\text{ADU}_{\text{c}}$) were extracted across the $900\text{--}1500\text{ eV}$ energy range measured at $-40\,^{\circ}\text{C}$ through the fitting procedure described above. The energy calibration curve was obtained by fitting the peak positions against the corresponding photon energies using a linear model, $\text{ADU}_{\text{c}} = g \cdot E + \text{offset}$, where $g$ represents the conversion gain and the offset accounts for residual baseline noise (which was found to be nearly zero). As illustrated in Fig.~\ref{fig:energy_calibration}, both the W17 (standard gain layer) and W9 (shallow gain layer) designs exhibit good linearity. The standard gain layer design demonstrates a significantly steeper calibration slope compared to the shallow gain layer, directly reflecting its higher multiplication factor.

\subsubsection{Temperature dependence of spectral response, multiplication gain and noise}

\begin{figure}[htbp]
    (a) \includegraphics[width=0.21\textwidth]{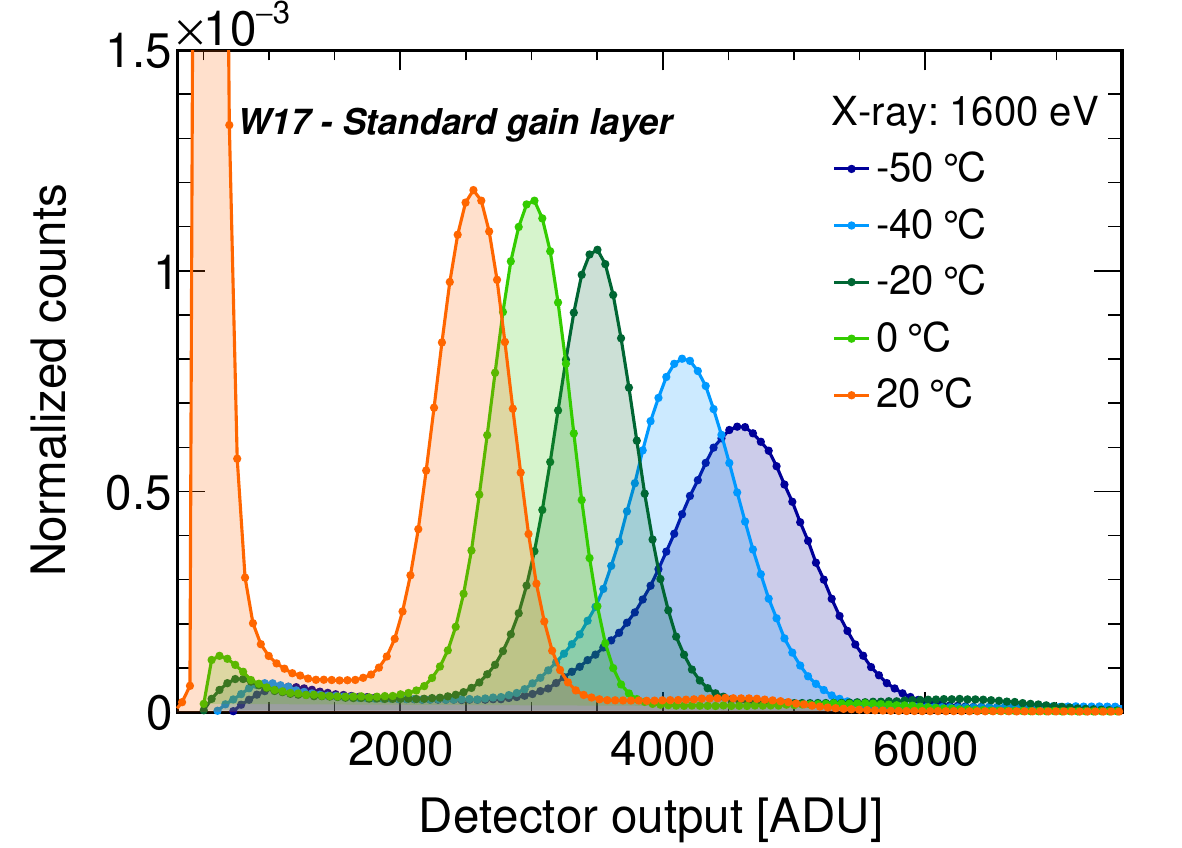}
    (b) \includegraphics[width=0.21\textwidth]{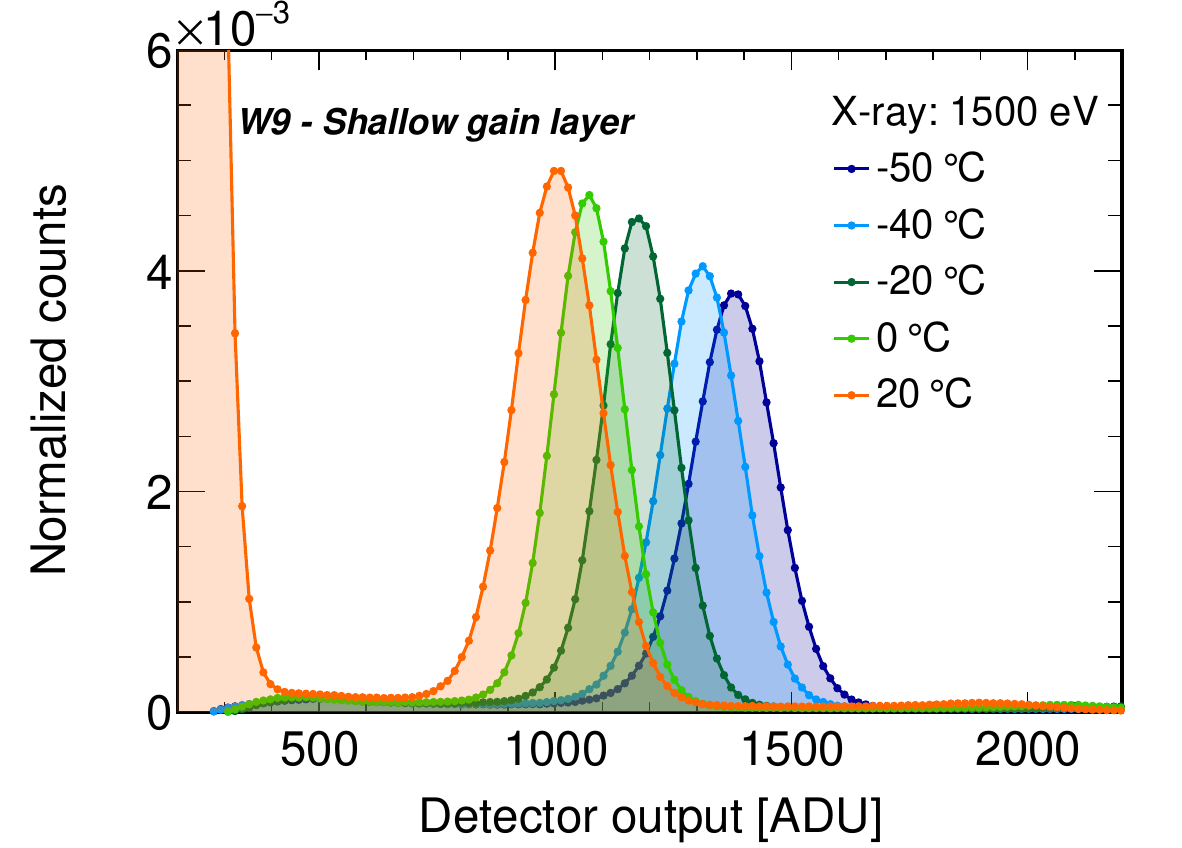}
    \caption{\label{fig_spectra_dependence_on_temperature} Spectra of the iLGAD sensors at temperatures ranging from \qty{-50}{\celsius} to \qty{20}{\celsius} for the (a) standard and (b) shallow gain layer designs. The spectra presented here are not calibrated to the incident X-ray energy.}
\end{figure}

To evaluate the temperature dependence of the sensor's performance, its spectral response, gain variation, and noise characteristics were analyzed. Fig.~\ref{fig_spectra_dependence_on_temperature} illustrates the spectral responses measured across a temperature range from \qty{-50}{\celsius} to \qty{20}{\celsius} at photon energies of \qty{1600}{\eV} (standard design) and \qty{1500}{\eV} (shallow design). Qualitatively, in both designs, the signal amplitude increases as the temperature decreases, driven by the enhanced impact-ionization coefficients of electrons and holes at low temperatures. However, this increase in the mean avalanche gain is accompanied by amplified statistical fluctuations inherent to the impact-ionization process. Consequently, the excess noise factor becomes the dominant contributor, which inevitably degrades the energy resolution at lower temperatures. 

\begin{figure}[ht]
    (a) \includegraphics[width=0.21\textwidth]{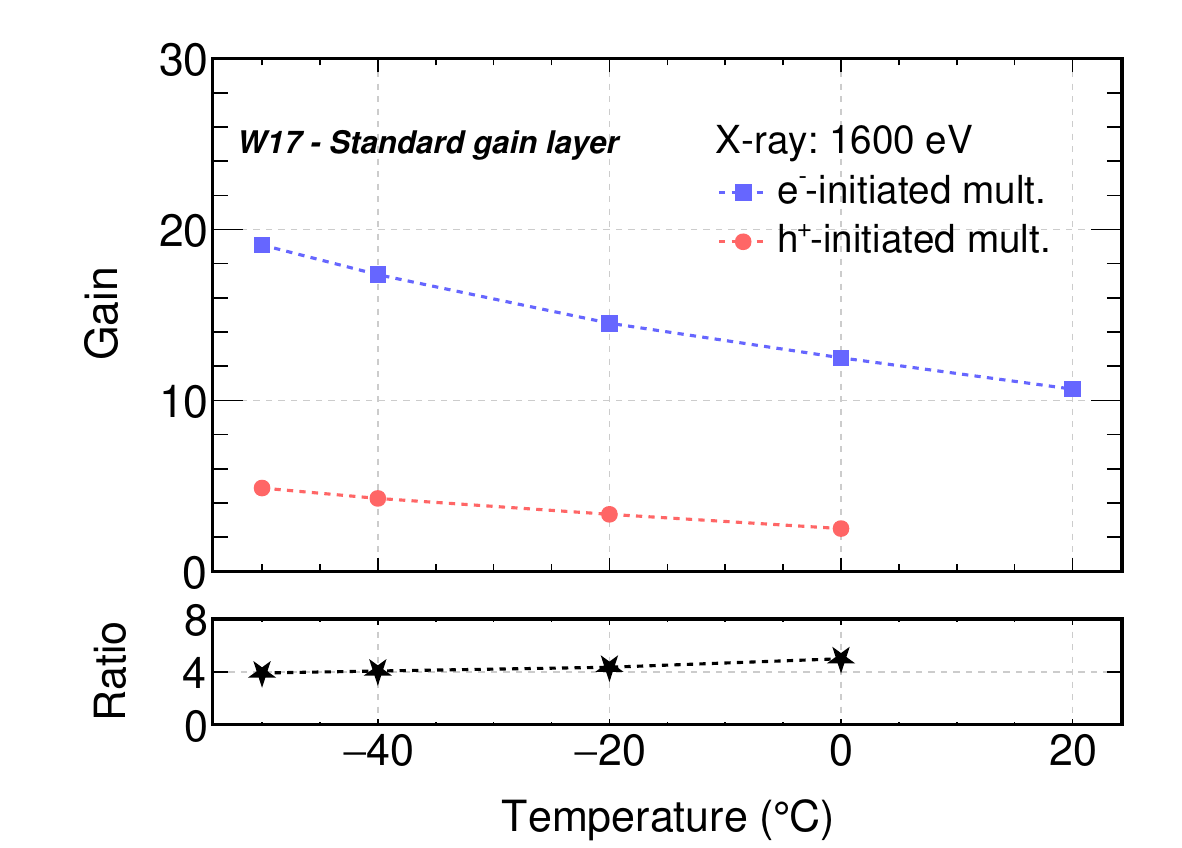}
    (b) \includegraphics[width=0.21\textwidth]{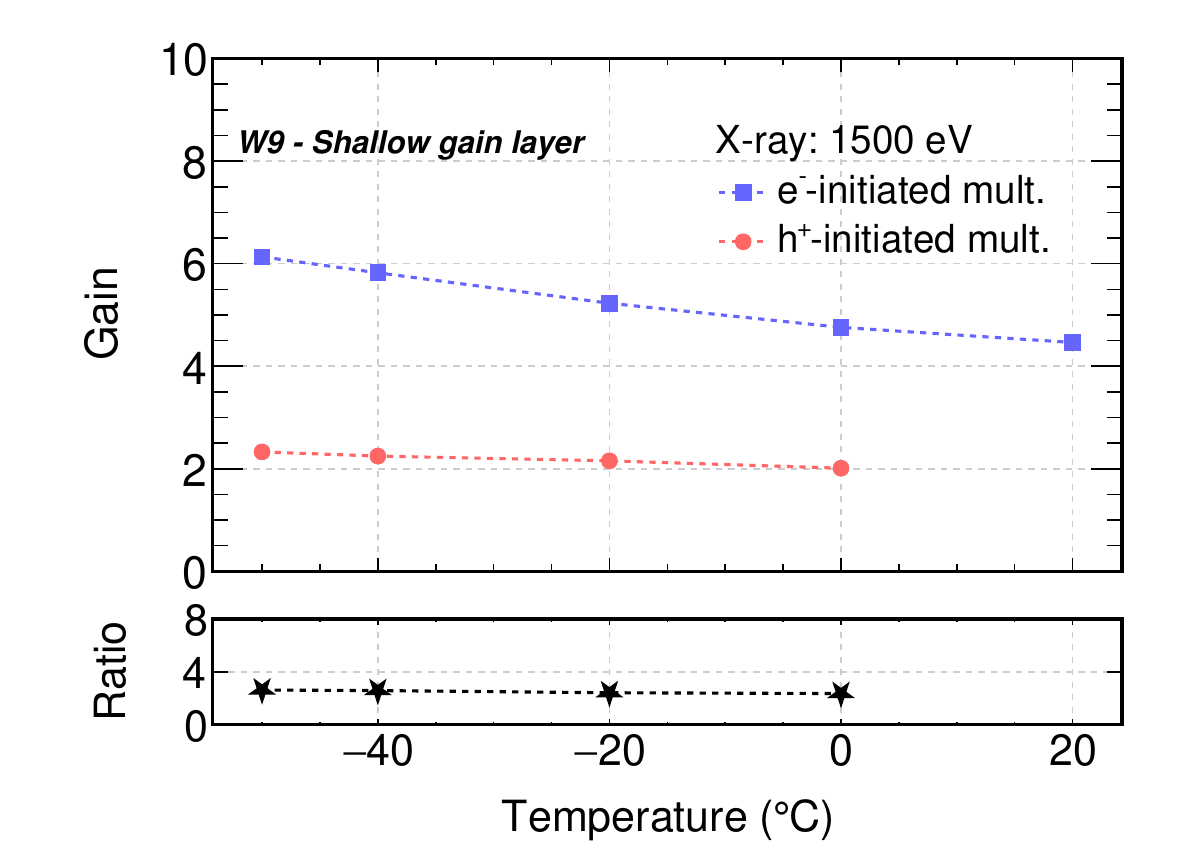}
    \caption{\label{fig_gain_vs_temperature} Multiplication factors of iLGAD sensors initiated by electrons and holes as a function of temperature for the standard (a) and shallow (b) gain layer designs.}
\end{figure}

To quantify this temperature-dependent gain variation, the electron- and hole-initiated multiplication factors were extracted from the spectra using the fitting procedure described in Sec.~\ref{sec:fitting}: The multiplication factor was calculated as the ratio between the extracted peak position of the iLGADs and that of the standard silicon sensor operating without gain. The results are presented in Fig.~\ref{fig_gain_vs_temperature}. At \qty{20}{\celsius}, the hole-initiated signal was buried in the high noise floor; thus, only the electron-initiated factor could be reliably determined. As the temperature decreased from \qty{20}{\celsius} to \qty{-50}{\celsius}, the electron-initiated factor in the standard design increased from $\sim$10.5 to 19.5, while the hole-initiated factor increased from $\sim$2.5 (measured at \qty{0}{\celsius}) to 5.0. The shallow design exhibited a weaker temperature dependence, with the electron-initiated factor increasing from $\sim$4.2 to 5.8, and the hole-initiated factor rising from $<1.8$ to $\sim$2.2 over their respective temperature ranges. At \qty{-40}{\celsius}, the ratio of the electron-to-hole multiplication factors was $\sim$4.0 for the standard design and $\sim$2.5 for the shallow design.

\begin{figure}[ht]
    \centering
    \includegraphics[width=0.4\textwidth]{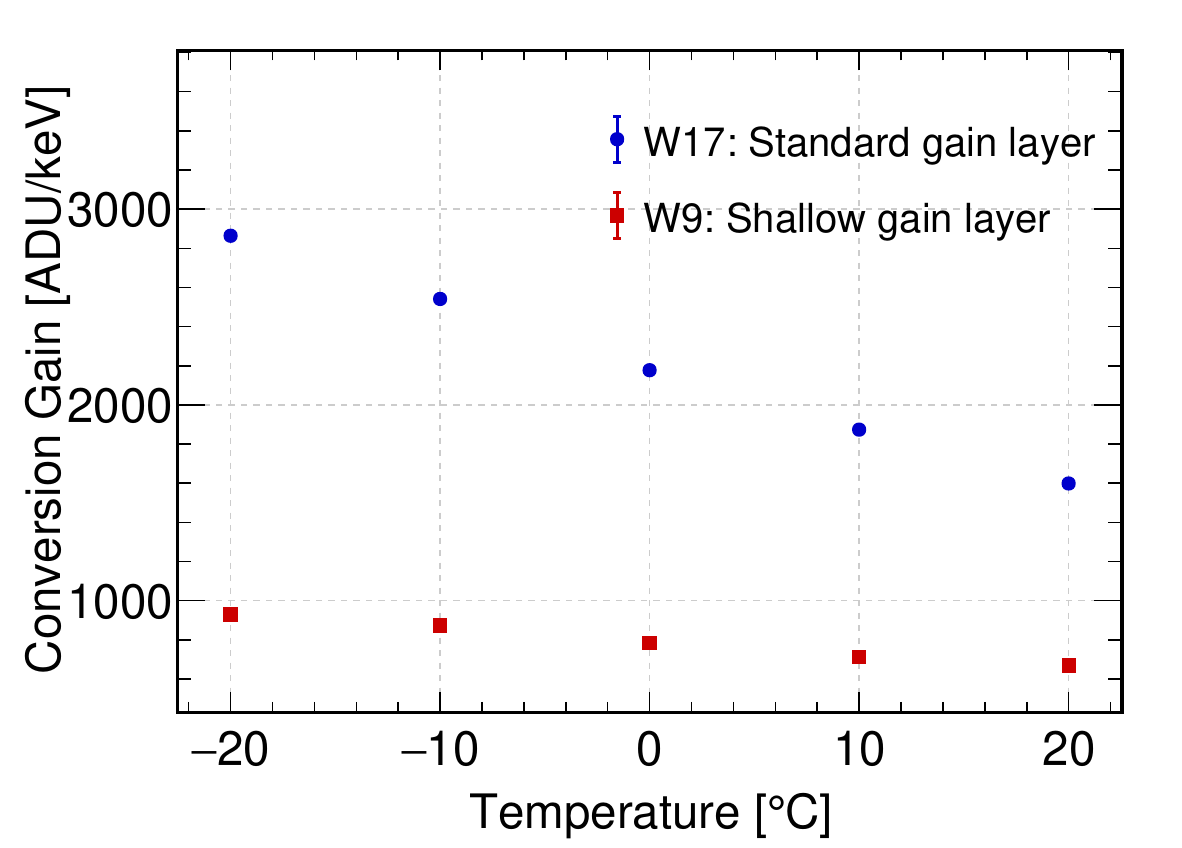}
    \caption{\label{fig:conversion_gain} Conversion gain at different temperatures for W9 and W17 iLGAD sensors.
}
\end{figure}

Alongside the gain, the electronic noise is fundamentally influenced by temperature and dictates the low-energy detection threshold. The effective equivalent noise charge (ENC), expressed in electrons ($e^-$), is calculated from the root-mean-square (r.m.s.) pedestal noise $\sigma$ (in ADU) and the conversion gain $G$ (in ADU/keV):

\begin{equation}
    \text{ENC} \; [e^-] = \frac{\sigma \; [\text{ADU}]}{G \; [\text{ADU/keV}]} \times \frac{1000 \; [\text{eV/keV}]}{w \; [\text{eV}/e^-]}
\end{equation}

\noindent where $w \approx \qty{3.62}{\eV}$ is the average energy required to create an electron-hole pair in silicon. The conversion gains at different temperatures are shown in Fig.~\ref{fig:conversion_gain}. These values were extracted following the energy calibration procedure described in Sec.~\ref{sec:energy_calibration}.

\begin{figure}[ht]
    \centering
    \includegraphics[width=0.4\textwidth]{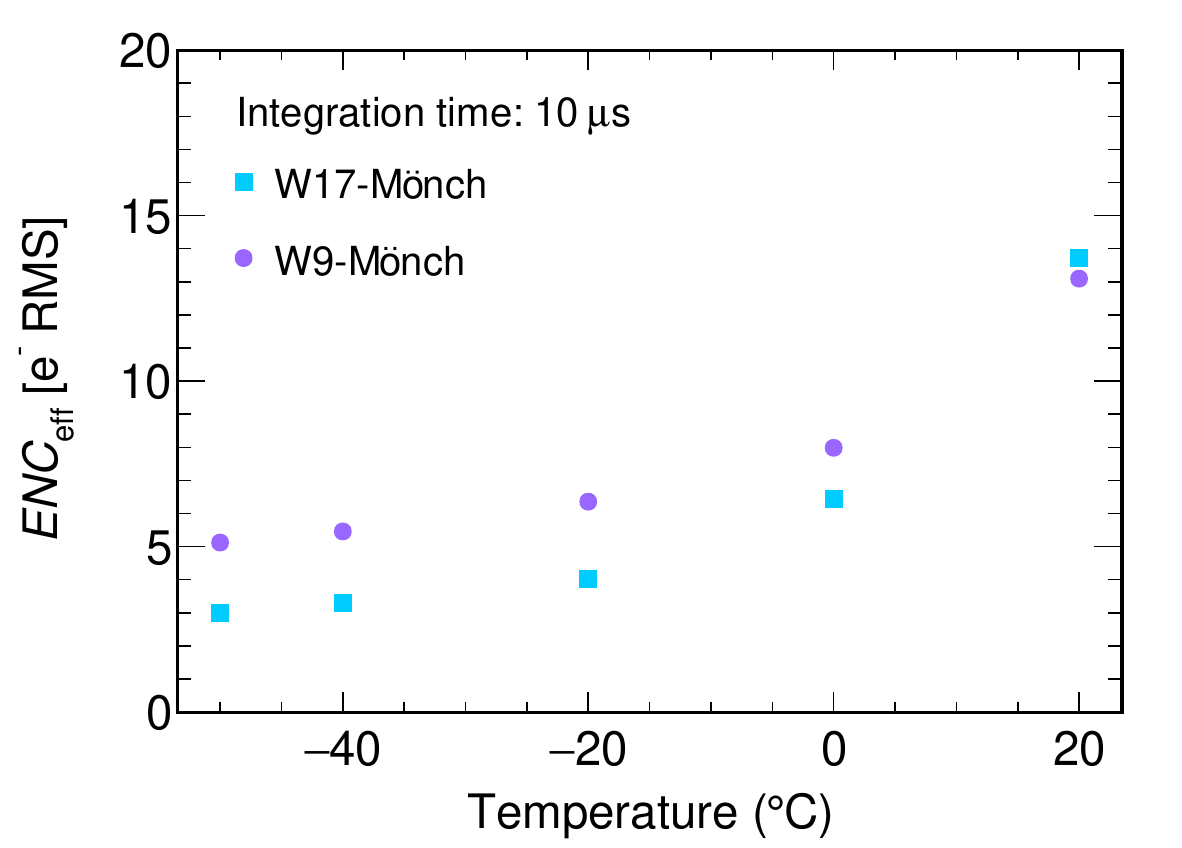}
    \caption{\label{fig_noise_temperature} Effective equivalent noise charge ($\text{ENC}_{\text{eff}}$) of the iLGAD-M\"onch module as a function of temperature for the standard (W17) and shallow (W9) gain layer designs.}
\end{figure}

As illustrated in Fig.~\ref{fig_noise_temperature}, the effective ENC decreases at lower temperatures owing to the suppressed leakage current and the increased multiplication gain. However, below \qty{-40}{\celsius}, the noise floor becomes dominated by the sensor's input capacitance, rendering further cooling ineffective. Therefore, \qty{-40}{\celsius} was selected as the optimal operating temperature to guarantee a sufficient SNR for soft X-ray measurements below \qty{1}{\keV}.

\subsubsection{Single-photon sensitivity and SNR}

\begin{figure}[htbp]
\centering
(a)\includegraphics[width=0.21\textwidth]{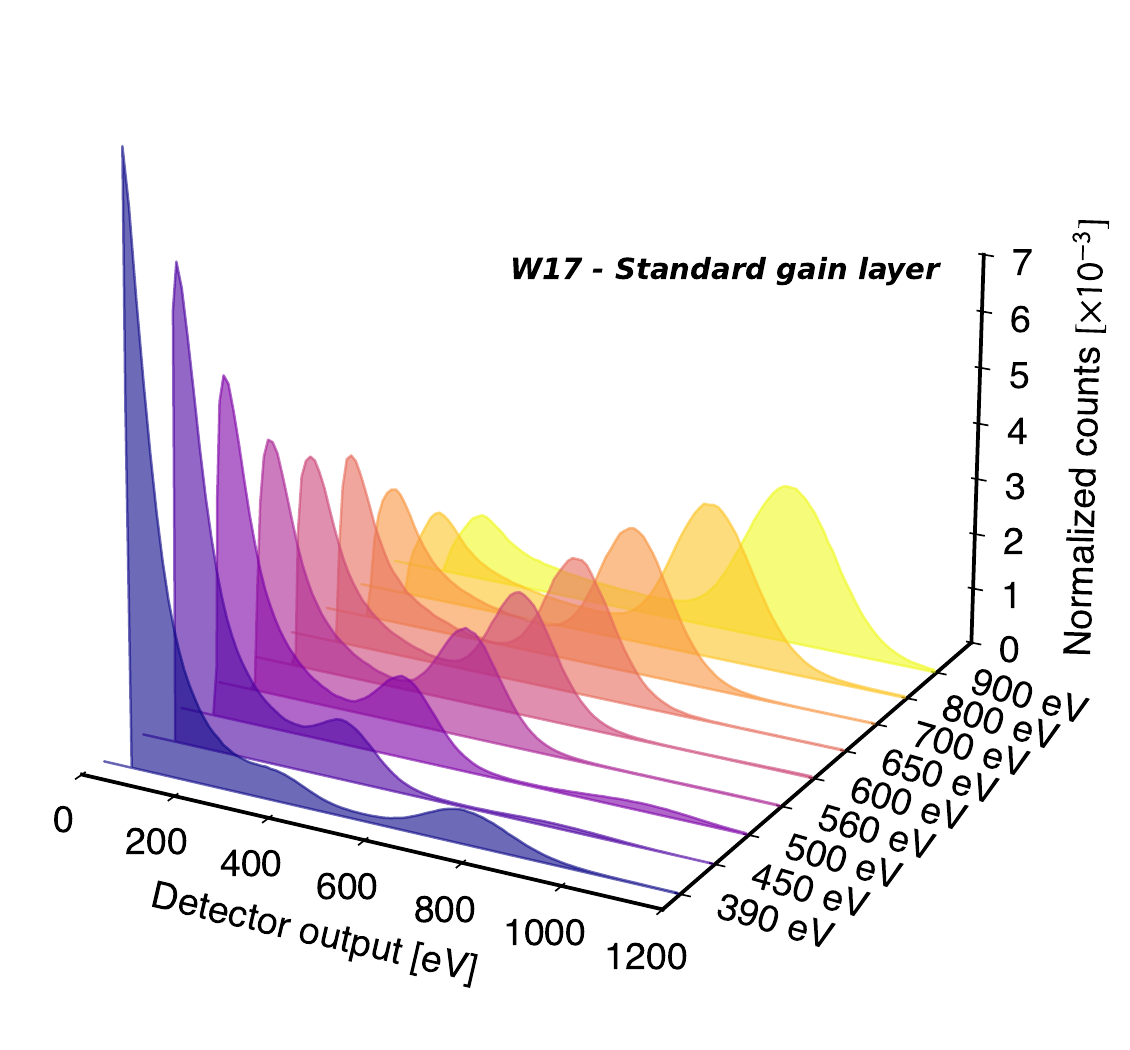}
(b)\includegraphics[width=0.21\textwidth]{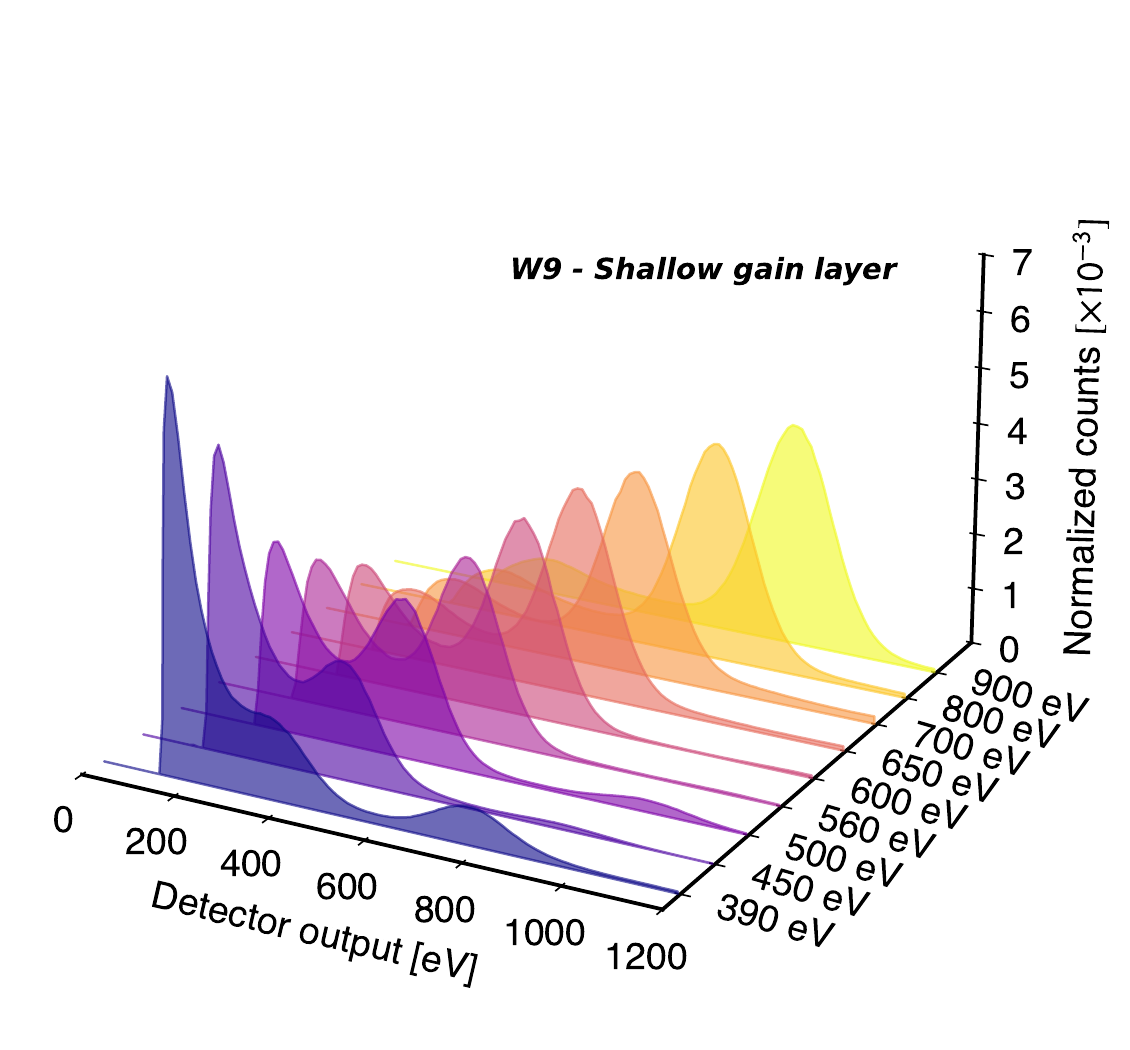}
\caption{Waterfall plots for standard (a) and shallow (b) iLGAD gain designs.}
\label{fig_spectral_response}
\end{figure}

To evaluate the single-photon detection capability of the iLGADs reported in this study, an energy scan from \qty{900}{\eV} down to \qty{390}{\eV} was performed at \qty{-40}{\celsius}.
Waterfall plots of the spectral response after calibration and noise-edge removal as a function of incident X-ray energy are presented in Fig.~\ref{fig_spectral_response}~(a) and (b) for the standard and shallow designs, respectively. 

In the shallow gain design, electron-initiated multiplication dominates above \qty{500}{\eV}, whereas in the standard design, it dominates only above \qty{700}{\eV}. This difference has key practical implications: while the standard design is suitable for single-photon counting at the Fe L-edge (\qty{703}{\eV}), it exhibits poor efficiency at the O K-edge (\qty{530}{\eV}) due to the high proportion of hole-initiated events. Consequently, the shallow design is preferred for O K-edge scientific applications. At the lowest tested energy of \qty{390}{\eV}, electron-initiated multiplication is completely suppressed, and the response is governed entirely by hole-initiated multiplication in both designs. Thus, for soft X-ray detection between the Ni K-edge (\qty{410}{\eV}) and the C K-edge (\qty{280}{\eV}), single-photon sensitivity depends entirely on the hole-initiated signal. This signal arises from absorption within the $\mathrm{n}^+$ implant; therefore, achieving effective detection in this energy regime demands high QE and CCE thanks to the previously developed TEW technology. Any charge loss due to carrier recombination in this heavily doped layer degrades the spectral performance, leading to peak broadening, low-energy tails, or a complete loss of single-photon sensitivity.

\begin{figure}[htbp]
\centering
(a)\includegraphics[width=0.21\textwidth]{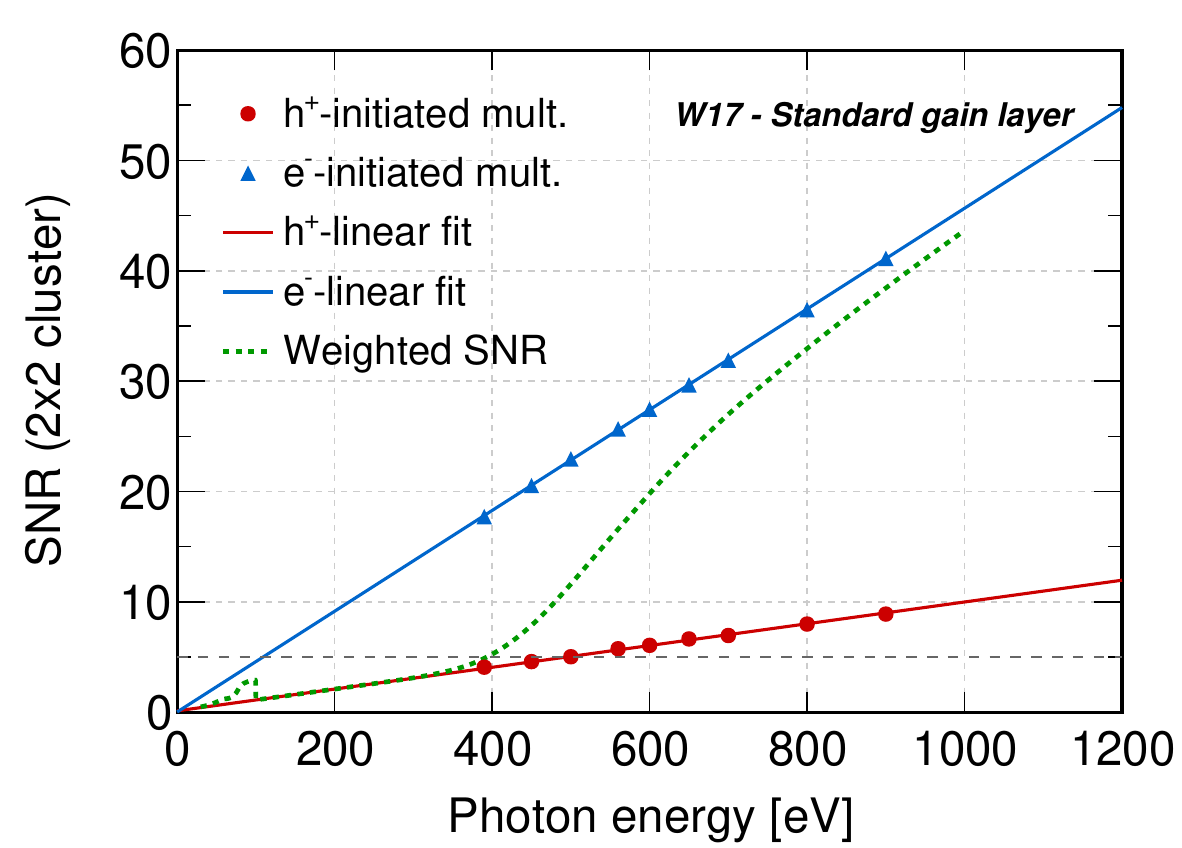}
(b)\includegraphics[width=0.21\textwidth]{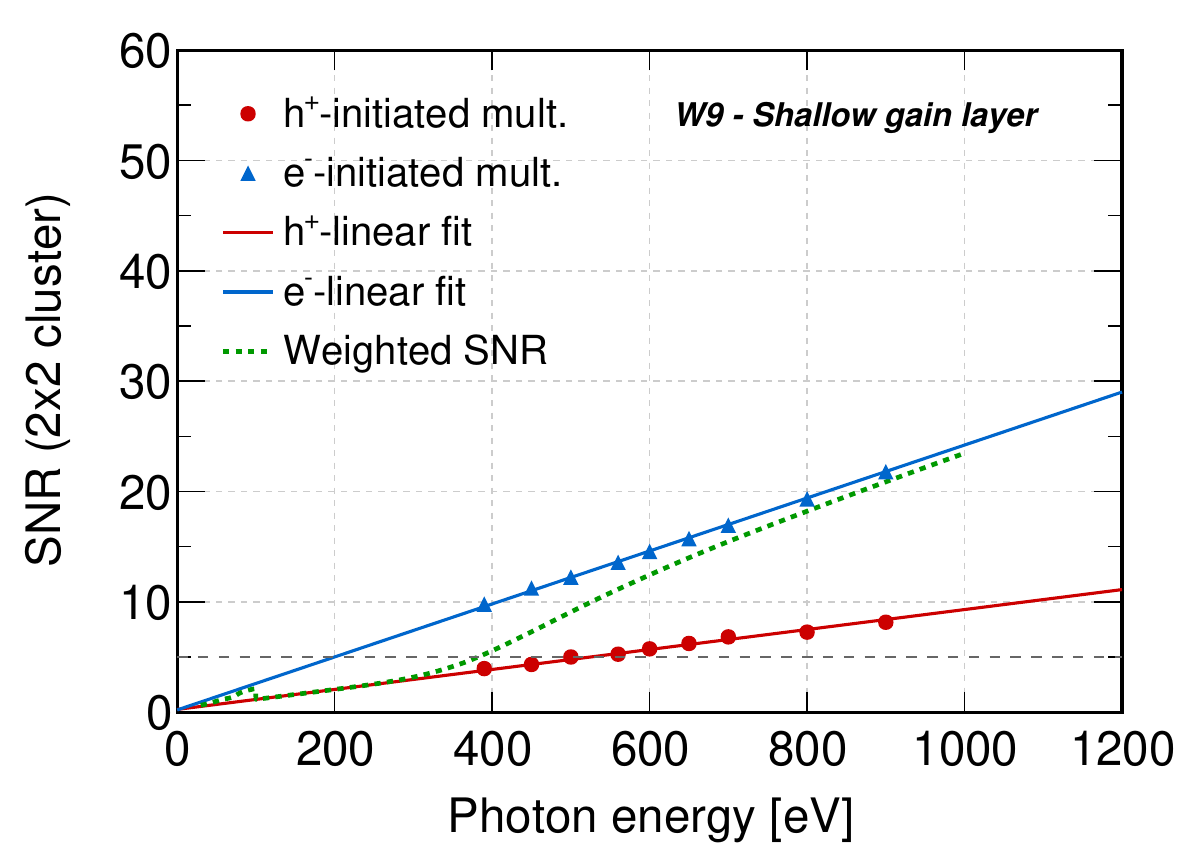}
\caption{Derived signal-to-noise ratios using the signals from electron- and hole-initiated multiplication for the two gain designs: (a) standard gain design, (b) shallow gain design. The green curve represents the combined weighted SNR.}
\label{fig_SNR}
\end{figure}

The SNRs of the electron- and hole-initiated multiplications were evaluated by extracting the two peak positions at each incident energy, utilizing the fitting procedure described in Sec.~\ref{sec:fitting}. These positions correspond to the mean signal pulse heights of the respective multiplication processes. The noise is defined as the r.m.s. of the detector output using dark images. The resulting electron- and hole-initiated SNRs for both designs are plotted in Fig.~\ref{fig_SNR}. Both SNRs scale linearly with the incident photon energy, indicating that the sensor response remains linear with no gain suppression in this energy regime and a good energy calibration discussed in the previous section. The weighted SNR (green curves in Fig.~\ref{fig_SNR}), which accounts for the relative fractions of the electron- and hole-initiated multiplication signals, reaches a value of 5 at $\sim$\qty{400}{\eV} for both designs, establishing the single-photon detection limit for the current iLGAD sensors. If the electron-initiated gain of the shallow design were increased to match that of the standard design, its inherently higher hole-initiated gain would yield a weighted SNR of 12.25. This indicates that an optimized shallow gain design could significantly improve soft X-ray sensitivity, potentially extending single-photon detection down to the C K-edge (\qty{280}{\eV}).

\subsubsection{Determination of the entrance window implant and gain layer depth}\label{sec3.3}

The variation in the fractional areas of the electron- and hole-initiated multiplication peaks at different energies reflects the photon absorption probability in different regions of the sensor. Following the spectral fitting, the spatial boundaries of the gain layer and the depletion boundries of the n$^{+}$-implant layer can be determined from the relative contributions of the three components in $f_{\text{total}}(x)$. According to the Beer–Lambert law, the probability of photon absorption at a depth $x$ is proportional to $\exp(-x/\lambda)$, where $\lambda$ is the photon attenuation length, obtained from the CXRO database~\cite{Henke1993, CXRO_optical_constants}.

The boundary separating the pure hole-initiated multiplication region from the co-multiplication region is denoted as $d_{\text{h}^{+}}$, which corresponds to the effective thickness of the $\mathrm{n}^+$ layer. Meanwhile, the boundary separating the co-multiplication region from the pure electron-initiated multiplication region is denoted as $d_{\text{e}^{-}}$, which defines the depth of the gain layer. 

The fraction of hole-initiated ($r_{\mathrm{h}}$) and electron-initiated ($r_{\mathrm{e}}$) multiplication components as a function of the photon attenuation length $\lambda$ can be expressed as:

\begin{subequations}
\label{eq:fractions}
\begin{align}
    r_{\mathrm{h}} &= 1 - \exp\left(-\frac{d_{\text{h}^{+}}}{\lambda}\right) \label{eq:rh} \\
    r_{\mathrm{e}} &= \exp\left(-\frac{d_{\text{e}^{-}}}{\lambda}\right)     \label{eq:re}
\end{align}
\end{subequations}

\noindent where $d_{\text{h}^{+}}$ and $d_{\text{e}^{-}}$ represent the characteristic depths extracted from the exponential fits. Here, $r_{\mathrm{h}}$ and $r_{\mathrm{e}}$ are calculated as the ratios of the integrated areas of $f_{\mathrm{h}}(x)$ and $f_{\mathrm{e}}(x)$ to the total integrated area of $f_{\mathrm{total}}(x)$, respectively.

\begin{figure}[htbp]
    (a) \includegraphics[width=0.21\textwidth]{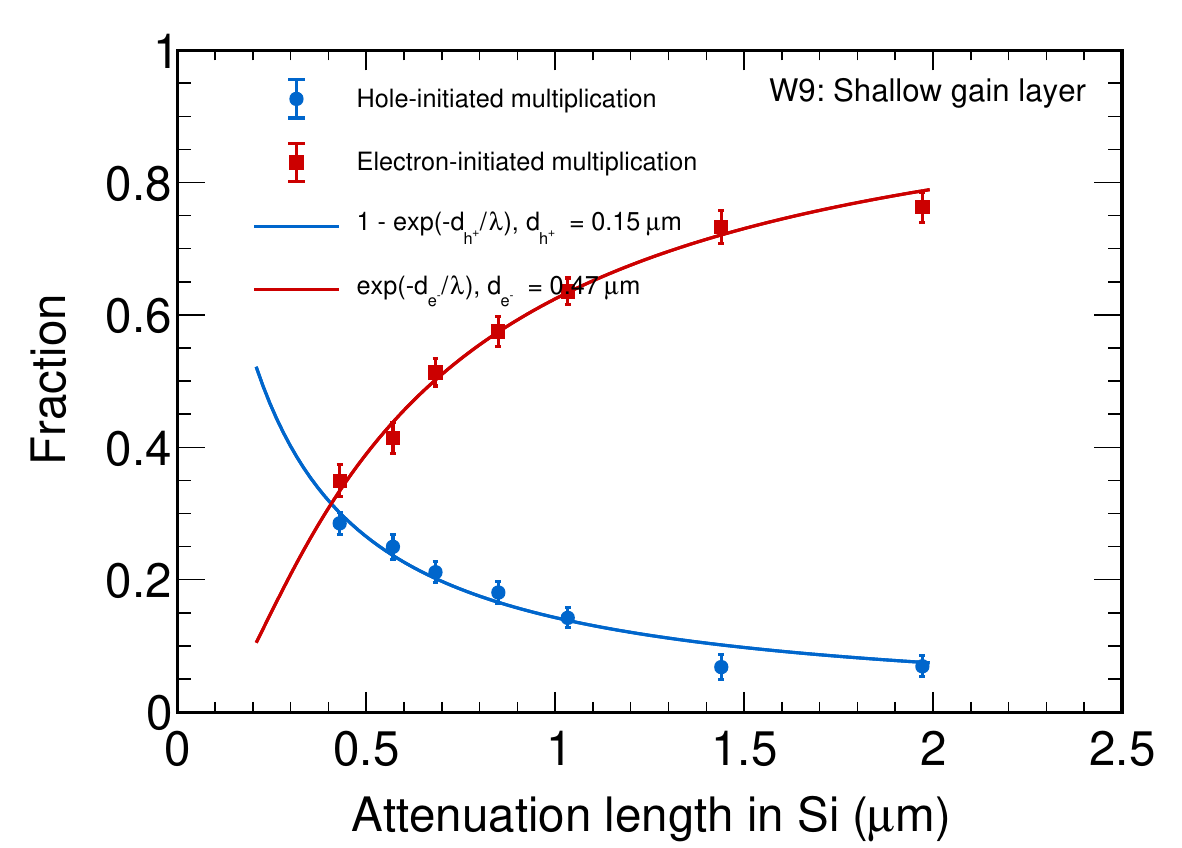}
    (b) \includegraphics[width=0.21\textwidth]{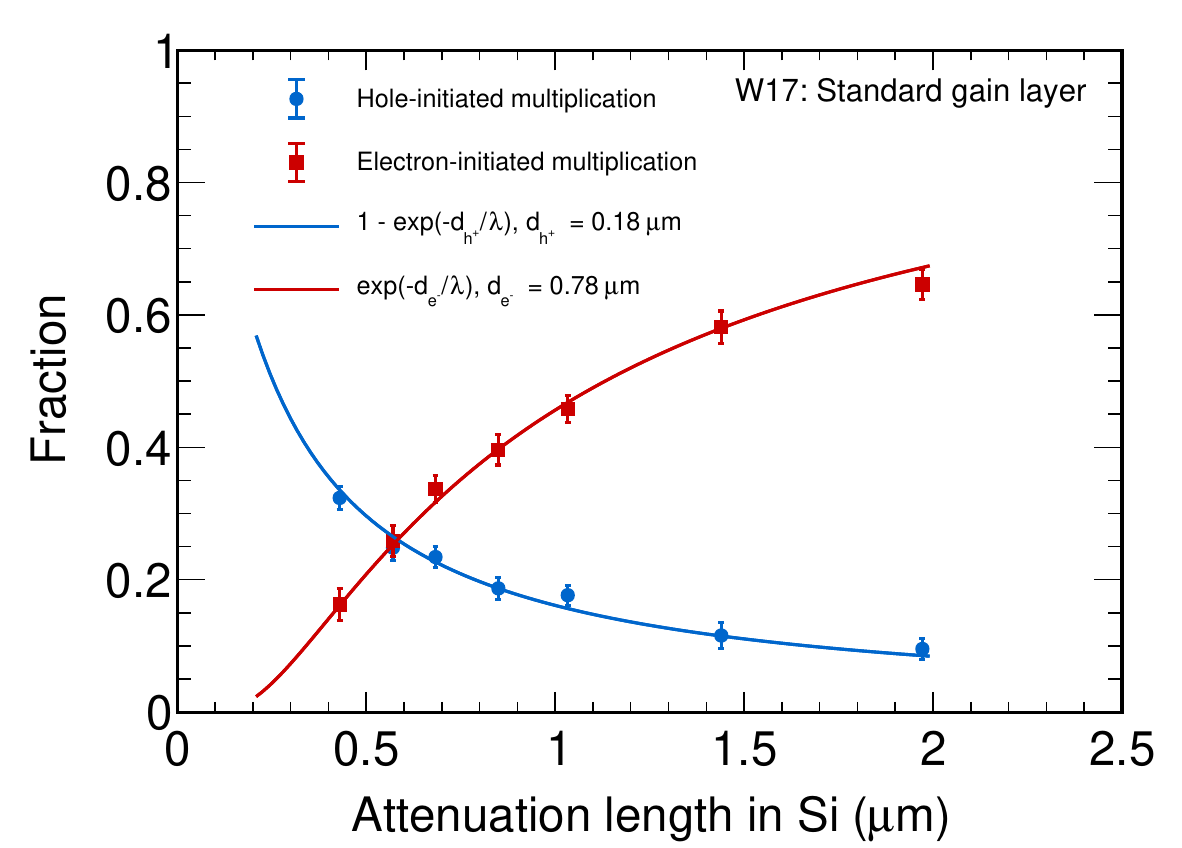}
    \caption{\label{fig:fraction_fit} Fraction of electron- and hole-initiated multiplication peaks as a function of X-ray attenuation length in silicon (derived from photon energies of 500 eV, 560 eV, 600 eV, 650 eV, 700 eV, 800 eV, and 900 eV) for the shallow (a) and standard (b) gain layer design. Symbols with error bars denote the data points, and solid lines represent the corresponding exponential fits. The extracted characteristic depths $d_{\text{e}^-}$ and $d_{\text{h}^+}$ are indicated in the legend.}
\end{figure}

As shown in Fig.\ref{fig:fraction_fit}, the characteristic parameters $d_{\text{h}^{+}}$ and $d_{\text{e}^{-}}$ for the standard and shallow gain layer LGAD designs are extracted by fitting the multiplication fraction versus silicon attenuation length. The two designs yield comparable $d_{\text{h}^{+}}$ values owing to their similar $\mathrm{n}^+$ doping profiles. In contrast, $d_{\text{e}^{-}}$ (representing the gain layer depth) for the shallow design is roughly half of that for the standard design, which agrees well with the TCAD doping profiles shown in Fig.\ref{fig:fig_iLGAD_opt}(c).


\section{Conclusion}\label{sec4}

In summary, this work provides a comprehensive study of the iLGAD sensors from our first R\&D batch integrated with a thin entrance window.
These developed iLGAD sensors featuring both standard and shallow gain-layer designs were fabricated and characterized using soft X-rays from a synchrotron radiation source. Due to the intrinsic structure of iLGADs and different photon absorption depths, the energy spectra exhibit a distinct double-peak response, which has been reported in detail in this work. The peak due to the hole-initiated charge multiplication has been observed thanks to the high charge collection efficiency for X-ray photons absorbed in the non-depleted n$^{+}$-implant layer after the implementation of a compatible thin entrance window process.
A composite fitting model was developed to quantify these measured spectra and extract the relative photon absorption ratios across different sensor regions. From these measurements, the depth of the non-depleted n$^{+}$-implant layer and the gain layer for the two designs can be determined, which is consistent with the design values obtained from the TCAD process simulation.
Measurements at different temperatures have been performed, indicating that cooling to -40~°C effectively suppresses leakage current while enhancing internal gain.
At the operating temperature of -40~°C, both iLGAD designs achieve an SNR of $\geq$5 down to 400~eV, demonstrating single photon sensitivity below the nitrogen K-edge. Our findings have proven that a shallow gain-layer design with enhanced electron-initiated multiplication is optimal for soft X-ray detection below 500~eV due to the smaller difference between the electron and hole-initiated multiplication factors. 

To extend single-photon sensitivity down to the carbon K-edge (280~eV), we utilize a shallow gain-layer architecture with a higher electron gain factor ($\sim$10, compared to the current value of 5) in our second-batch iLGAD sensors, which is expected to yield an SNR of $\geq$5 at 280~eV. Overall, these advancements in hybrid photon detectors pave the way for high-efficiency, high-speed, and wide-dynamic-range soft X-ray imaging and spectroscopy at next-generation light sources.

\begin{acknowledgments}
We acknowledge MAX IV Laboratory for time on Beamline SoftiMAX. We sincerely thank Karina Th\aa{}nell and Igor B\'enyk from the SoftiMAX beamline for their fruitful discussions and invaluable assistance during the measurements.

\end{acknowledgments}
\section*{AUTHOR DECLARATIONS}
\subsection*{Conflict of Interest}
The authors have no conflicts of interest to disclose.

\section*{DATA AVAILABILITY}
The data that support the findings of this study are available from the corresponding author upon reasonable request.





\bibliography{Soft_Xray_iLGAD_V0}

\end{document}